\documentclass{aastex6}
\usepackage{mathtools}
\usepackage{amsmath}
\usepackage{float}
\usepackage{longtable}
\usepackage{soul}

\DeclareMathOperator{\sinc}{sinc}
\fullcollaborationName{The Friends of AASTeX Collaboration}

\begin{document}


\title{ARMADA I: Triple Companions Detected in B-Type Binaries $\alpha$ Del and $\nu$ Gem}


\author{Tyler Gardner\altaffilmark{1}, John D. Monnier\altaffilmark{1}, Francis C. Fekel\altaffilmark{2}, Gail Schaefer\altaffilmark{3}, Keith J.C. Johnson\altaffilmark{1}, Jean-Baptiste Le Bouquin\altaffilmark{4}, Stefan Kraus\altaffilmark{5}, Narsireddy Anugu\altaffilmark{6}, Benjamin R. Setterholm\altaffilmark{1}, Aaron Labdon\altaffilmark{5}, Claire L. Davies\altaffilmark{5}, Cyprien Lanthermann\altaffilmark{7}, Jacob Ennis\altaffilmark{1}, Michael Ireland\altaffilmark{8}, Kaitlin M. Kratter\altaffilmark{6}, Theo Ten Brummelaar\altaffilmark{3}, Judit Sturmann\altaffilmark{3}, Laszlo Sturmann\altaffilmark{3}, Chris Farrington\altaffilmark{3}, Douglas R. Gies\altaffilmark{3}, Robert Klement\altaffilmark{3}, Fred C. Adams\altaffilmark{1,10}}

\altaffiltext{1}{Astronomy Department, University of Michigan, Ann Arbor, MI 48109, USA}
\altaffiltext{2}{Center of Excellence in Information Systems, Tennessee State University, Nashville, TN 37209, USA}
\altaffiltext{3}{The CHARA Array of Georgia State University, Mount Wilson Observatory, Mount Wilson, CA 91203, USA}
\altaffiltext{4}{Institut de Planetologie et d'Astrophysique de Grenoble, Grenoble 38058, France}
\altaffiltext{5}{Astrophysics Group, Department of Physics \& Astronomy, University of Exeter, Stocker Road, Exeter, EX4 4QL, UK}
\altaffiltext{6}{Steward Observatory, Department of Astronomy, University of Arizona, 933 N. Cherry Ave, Tucson, AZ, 85721, USA}
\altaffiltext{7}{Institute of Astronomy, KU Leuven, Celestijnenlaan200D, 3001, Leuven, Belgium}
\altaffiltext{8}{Research School of Astronomy \& Astrophysics, Australian National University, Canberra ACT 2611, Australia}
\altaffiltext{9}{Department of Astronomy and Steward Observatory, Univ. of Arizona, 933 N Cherry Ave, Tucson, AZ, 85721}
\altaffiltext{10}{Physics Department, University of Michigan, Ann Arbor, MI 48109, USA}

\begin{abstract}
Ground-based optical long-baseline interferometry has the power to measure the orbits of close binary systems at $\sim$10 micro-arcsecond precision. This precision makes it possible to detect ``wobbles" in the binary motion due to the gravitational pull from additional short period companions. We started the ARrangement for Micro-Arcsecond Differential Astrometry (ARMADA) survey with the MIRC-X instrument at the CHARA array for the purpose of detecting giant planets and stellar companions orbiting individual stars in binary systems. We describe our observations for the survey, and introduce the wavelength calibration scheme that delivers precision at the tens of micro-arcseconds level for $<$0.2 arcsecond binaries. We test our instrument performance on a known triple system $\kappa$ Peg, and show that our survey is delivering a factor of 10 better precision than previous similar surveys. We present astrometric detections of tertiary components to two B-type binaries: a 30-day companion to $\alpha$ Del, and a 50-day companion to $\nu$ Gem. We also collected radial velocity data for $\alpha$ Del with the Tennessee State University Automated Spectroscopic Telescope at Fairborn Observatory. We are able to measure the orbits and masses of all three components in these systems. We find that the previously published RV orbit for the inner pair of $\nu$ Gem is not consistent with our visual orbit. The precision achieved for these orbits suggests that our ARMADA survey will be successful at discovering new compact triple systems to A/B-type binary systems, leading to better statistics of hierarchical system architectures and formation history.
\end{abstract}

\keywords{astrometry, binaries: close, technique: interferometry}



\section{Introduction} 
\label{sec:intro}
Binary systems are of vital importance for understanding the physical properties of stars by providing the opportunity to directly measure stellar masses. Precision visual binary orbits can be combined with single or double-lined radial velocity (RV) data to solve for the physical parameters of the system (i.e., masses, physical orbital elements, mutual inclinations for tertiaries, position on HR diagram). Along with the importance of providing mass and stellar evolution information, visual binary orbits at the micro-arcsecond ($\mu$as) level of precision can open up the possibility to search for additional inner companions to the outer binary system. When a binary orbit is monitored over a long enough time scale, unseen inner companions will impart an additional ``wobble" motion on top of the binary motion as the inner pair orbits its center-of-mass. Once instrumental precision reaches the $\sim$10$\mu$as level, this method can be used to detect companions down to the regime of Jupiter-mass planets. The power of this method opens up exciting new areas of parameter space in which to search for exoplanets and stellar companions. 

For mass ratios $>$0.1 for the inner pair, tertiary occurrence rate is expected to be roughly 30\% for A/B-type binaries with semi-major axes of the outer pair 10-100 astronomical units (au) \citep{tokovinin2014, moe2017}. Multiplicity surveys are rather incomplete however for inner companions with M$_{\rm{companion}}$ / M$_{\rm{star}}$ $<$ 0.4 in 0.5-5 au orbits (see Fig 1 of \citealt{moe2017}), as there are few observation methods which can fill this gap. We know that in general triple systems appear to be more common for higher mass stars \citep{maz2019}, making high mass stars a potentially fertile regime for detecting inner subsystems to the binary. In the planetary regime, the frequency of $\sim$1 au giant planets around ``hot" A/B-type main sequence stars is very uncertain due to the difficulty of RV surveys for these stars with weak and broad spectral lines. Surveys of evolved stars suggest an increase in $\sim$1 au giant planets for massive stars compared with solar mass stars \citep{johnson2010a,bowler2010,ghezzi2018}. The true progenitor masses of these surveys are often disputed however (e.g. \citealt{lloyd2011}). Direct imaging work is also seemingly consistent with a top-heavy distribution in stellar mass for planet frequency \citep{nielsen2019,vigan2020}, though some high precision RV work has not seen the same increase in au giant planet frequency with mass \citep{borgniet2019}. A method that could search for companions down to the planetary regime around main-sequence A/B-type stars would improve stellar multiplicity statistics, as well as help resolve some of the disputes about the affects of stellar mass on giant planet frequency.

Optical long-baseline interferometry has long been useful for delivering the visual orbits of binaries (see \citealt{bonneau2014} for a review on the importance of interferometric measurements of binaries). As instrumental precision advances, visual binary orbits can be constrained more tightly leading to better measurements of masses, radii, and evolutionary stage. \citet{Muterspaugh2010} demonstrated with the Palomar Testbed Interferometer that high precision differential astrometry of binary systems can be used to detect additional companions of binary systems down to the planetary regime. Current instruments have improved since that survey, and optical long-baseline interferometers can now routinely deliver astrometric measurements down to the $\sim$10$\mu$as precision level (e.g. GRAVITY at VLTI: \citealt{gravity2017}; MIRC/MIRC-X at CHARA: \citealt{gardner2018}, \citealt{schaefer2016}). A dedicated interferometric survey of wide ``hot star" binaries thus has the power to reveal companions down to the mass regime of $\sim$au giant planets via differential astrometry. We started the ARrangement for Micro-Arcsecond Differential Astrometry (ARMADA) survey with the primary goal of detecting $\sim$au giant exoplanets in A/B-type binary systems. In this paper we present data taken for the ARMADA survey with the MIRC-X instrument \citep{anugu2020}, which is a recent upgrade from the Michigan Infra-Red Combiner (MIRC) \citep{Monnier2006}. MIRC-X is a H-band combiner of six 1-meter telesopes at the Georgia State University Center for High Angular Resoloution Astronomy (CHARA) Array. Our ultimate goal with ARMADA is to reveal $\sim$1 au circumstellar giant planets in binary systems (i.e., planets that orbit an individual star of the binary pair), though a large number of epochs are required to detect these small signals. Compact triple systems, where the previously unseen third companion is of stellar mass, are relatively easy to detect with our ARMADA survey since they induce astrometric ``wobbles" at the 1000s of micro-arcsecond level. In this paper we present three compact triple orbits detected with ARMADA data: 1) a known F-Type compact triple system $\kappa$ Peg as a validation of our methods, 2) a new compact triple in the B-type binary system $\alpha$ Del, and 3) the first astrometric detection of a tertiary in the system of B-type binary $\nu$ Gem.

In section \ref{sec:observations} we describe our observations and data reduction methods for ARMADA, along with our wavelength calibration scheme. Section \ref{sec:orbitfitting} outlines our orbit fitting models. Section \ref{sec:kappeg} shows ARMADA results on a known compact triple system as a test to our observation and calibration methods. In section \ref{sec:alpdel} and \ref{sec:nugem} we present our new tertiary companions to $\alpha$ Del and $\nu$ Gem, along with the best fit orbits. Section \ref{sec:formation} describes the formation history constraints from these orbits. We give concluding remarks and prospects for future results in section \ref{sec:conclusion}.

\section{Observations and Data Reduction} 
\label{sec:observations}
\subsection{MIRC-X at the CHARA Array}
Data for the ARMADA survey are taken in $H$-band with the Michigan InfraRed Combiner-Exeter (MIRC-X) instrument at the CHARA Array. The CHARA Array is the optical/near-IR interferometer with the longest baselines in the world \citep{Brummelaar2005}. MIRC-X combines all six telescopes available at CHARA with baselines up to 330 m. The original MIRC instrument is described in detail by \citet{Monnier2006}. In 2017 July the detector and combiner were upgraded to MIRC-X \citep{anugu2020}. Epochs for the ARMADA survey are all taken in grism (R$\sim$190) mode, allowing us to detect components out to $\sim$200 mas with the larger interferometric field-of-view. Observational details and calibrators used for MIRC-X observations are displayed in Tables \ref{mircx_obs} and \ref{mirc_cals}. Unless otherwise noted in the table, epochs were collected with all six telescopes.

\begin{table} [H]
\begin{center}
\caption{Log of MIRC-X interferometric observations.\label{mircx_obs}}
\begin{tabular}{llccc}
\tableline
\tableline
\textsc{UT} date & Target & No. of 60-sec averages & Calibrators\tablenotemark{a} & Notes\\
\tableline
2017 Sep 28 & $\nu$ Gem & 8 & HD886 & \\
2017 Sep 30 & $\nu$ Gem & 10 & HD219080 & \\
2018 Jul 19 & $\alpha$ Del & 8 & HD176437 & \\
2018 Aug 21 & $\alpha$ Del & 11 & HD176437 & \\
2018 Sep 19 & $\alpha$ Del & 13 & HD886 & \\
 & $\kappa$ Peg & 20 & HD886 & \\
2018 Sep 20 & $\nu$ Gem & 9 & HD886 & \\
 & $\kappa$ Peg & 18 & HD886 & \\
2018 Nov 21 & $\nu$ Gem & 9 & HD886 & 5-telescopes (no S1) \\
 & $\kappa$ Peg & 8 & HD886 & 5-telescopes (no S1) \\
2018 Dec 04 & $\nu$ Gem & 14 & HD37202 & \\
2019 Jun 01 & $\alpha$ Del & 14 & HD176437, STS & \\
 & $\kappa$ Peg & 9 & HD176437, STS & \\
2019 Jun 03 & $\alpha$ Del & 9 & HD161868, STS & \\
2019 Jul 29 & $\alpha$ Del & 15 & HD176437, STS & \\
2019 Jul 30 & $\alpha$ Del & 10 & HD219080, STS & \\
 & $\kappa$ Peg & 10 & HD219080, STS & \\
2019 Jul 31 & $\alpha$ Del & 9 & HD886, STS & \\
 & $\kappa$ Peg & 15 & HD886, STS & \\
2019 Aug 01 & $\alpha$ Del & 10 & HD177756, STS & \\
2019 Aug 06 & $\alpha$ Del & 10 & HD886, STS & \\
 & $\kappa$ Peg & 10 & HD886, STS & \\
2019 Aug 08 & $\alpha$ Del & 10 & HD886, STS & \\
 & $\kappa$ Peg & 6 & HD886, STS & \\
2019 Sep 08 & $\alpha$ Del & 8 & HD886, STS & \\
 & $\nu$ Gem & 19 & HD886, STS & \\
2019 Oct 13 & $\nu$ Gem & 15 & HD219080, STS & \\
2019 Nov 11 & $\nu$ Gem & 10 & HD219080, STS & \\
2019 Nov 12 & $\alpha$ Del & 10 & HD219080, STS & 5-telescopes (no E1) \\
\tableline
\end{tabular}
\tablenotetext{a}{Refer to Table~\ref{mirc_cals} for details of the calibrators used. STS is our internal calibration source, implemented in 2019May.}
\end{center}
\end{table}

\begin{table} [H]
\begin{center}
\caption{On-Sky Calibrators\tablenotemark{1}\label{mirc_cals}}
\begin{tabular}{lccc}
\tableline
\tableline
HD & Sp. type & $H$ (mag) & $\theta_\mathrm{UD,H-band}$ (mas) \\
\tableline
886 & B2 IV & 3.43 & $0.41 \pm 0.03$ \\
176437 & B9 III & 3.19 & $0.72 \pm 0.08$ \\
161868 & A1 V & 3.64 & $0.616 \pm 0.05$ \\
219080 & F1 V & 3.76 & $0.693 \pm 0.07$ \\
177756 & B9 V & 3.64 & $0.597 \pm 0.06$ \\
37202 & B1 V & 3.05 & $0.519 \pm 0.05$ \\
\tableline
\end{tabular}
\tablenotetext{1}{Calibrator information was gathered from the JMMC SearchCal tool \citep{bonneau2006}}
\end{center}
\end{table}

The MIRC-X combiner measures visibilities, differential phase, and closure phase of our targets. Normally, one employs frequent observations of nearby calibrator stars to measure visibility loss due to time-variable factors such as atmospheric coherence time, vibrations, differential dispersion, and birefringence in the beam train. For ARMADA our main interest is differential astrometry between two components of a binary system within the interferometric field-of-view (both components generally unresolved). Since closure phase is immune to atmospheric effects, and extra dispersion in differential phase can be fit with a polynomial, we are able to observe for ARMADA without the use of traditional calibrators by fitting to closure and differential phase. This is crucial for ARMADA operations, as we are able to spend most of the night observing targets rather than spending time on calibrator sources. We still use sparse on-sky calibrators for our wavelength calibration (described in section \ref{sec:observations-wavelength}) as well as for a rough calibration of visibilities. In May 2019, the MIRC-X team implemented an internal light source called Six Telescope Simulator (STS, described in \citealt{anugu2020}). This higher SNR source is now used for wavelength calibration instead of the on-sky calibrator in dates following its implementation. We used the MIRC-X data pipeline (version 1.3.3) to produce OIFITS files for each night, described in \citet{anugu2020}. This pipeline and its documentation is maintained on Gitlab\footnote{\url{https://gitlab.chara.gsu.edu/lebouquj/mircx_pipeline}}. These nights were reduced with the ``spectral-differential" method of the MIRC-X pipeline for computing differential phase. This method first removes the group delay from the raw phase, and then computes differential phase as the phase(i+1) - phase(i) where i, i+1 are neighboring wavelength channels. We reduced most of our data with the number of coherent integration frames (ncoh) of 10, oifits max integration time of 60 seconds, and bispectrum bias correction applied. Since the companion to $\kappa$ Peg is near the edge of the interferometric field-of-view ($>$150 mas for latter epochs), its binary phase signal is varying faster on some baselines than the 60-second integration time. Hence we reduce the $\kappa$ Peg epochs from 2019 June onward with an oifits max integration time of 10 seconds. Since this object is bright, the high SNR allows us to combine fewer frames into a measurement of phase. Data taken in 2018 September showed signs of a vibration present within our combiner, with a quick loss of coherence reported by the pipeline. This forced us to reduce any 2018 September data with a lower number of ncoh (3 frames, instead of 10).

\subsection{Fitting Binary Star Differential Astrometry}
For each MIRC-X night we fit to the following binary model of complex visibility, $V$:

\begin{equation}
    V = \frac{V_1 + \Gamma f V_2 e^{-2\pi i (u\alpha+v\delta)}}{1+f}.
\label{vis_eqn}
\end{equation}
The free parameters for this binary model include a uniform disk for the primary and secondary to form visibilities $V_1$ and $V_2$; a binary separation in right ascension (R.A.) and declination (DEC) -- ($\alpha$, $\delta$); a monochromatic flux ratio between the two components $f$; as well as a bandwidth smearing parameter $b = 1/R$, where $R$ is the resolution of the disperser and $\Gamma = \sinc[b(u\alpha+v\delta)]$. The location on the uv-plane is denoted by parameters $u$ and $v$.

Since we do not use the standard CAL-SCI sequence of observing, our squared visibilities are poorly calibrated. We thus use the closure phase and differential phase observables to fit our binary positions for each epoch. To inform our uniform diameter (UD) values in our fits, we choose a well calibrated epoch for each target (i.e., a dataset that is taken near to the nightly calibrator star). This choice has little affect on the differential astrometry, since these targets are all mostly unresolved. For $\alpha$ Del we use the 2019Jul29 epoch (calibrator HD 176437) to fit for uniform diameters. We see no $\chi^2$ improvement in the fits by letting the diameters vary, and the two companions of this triple system prefer diameter values close to point sources. Hence we fix all three diameters to $UD=0.5$ mas, which is near the resolution limit in H-band. \citet{gordon2019} measured a uniform diameter of $0.407\pm0.022$ mas for the A component of $\alpha$ Del, consistent with our fixed values. $\nu$ Gem also showed no improvement in the fits by letting the UDs vary for the night of 2017Sep30 (calibrator HD219080). Hence we again fix all three diameter to $UD=0.5$ mas. For the well-calibrated 2019Aug06 epoch of $\kappa$ Peg (calibrator HD 886), we measure diameters of $UD_1=0.730\pm0.002$ mas and $UD_2=0.733\pm0.003$ mas. We then fix the uniform diameters for all $\kappa$ Peg epochs to $UD=0.7$ mas. The flux ratio values we report are the fitted values from these well-calibrated epochs.

In the case of $\alpha$ Del and some epochs of $\nu$ Gem we also detect flux from a third component in the system, as judged by significant residuals in a fit of the binary model to the observables. In this case, an extra component is added to our complex visibility model:

\begin{equation}
    V = f_1 V_1 + \Gamma_{12} f_2 V_2 e^{-2\pi i (u\alpha_{12}+v\delta_{12})} + \Gamma_{13} f_3 V_3 e^{-2\pi i (u\alpha_{13}+v\delta_{13})}.
\end{equation}
The symbols have their same meaning as in the binary model, except now in our notation $f_1, f_2, f_3$ are the flux contributions from each component and $f_1 + f_2 + f_3 = 1$. Note there are now two differential positions in R.A. and DEC - the primary to the secondary ($\alpha_{12}$, $\delta_{12}$), and the primary to the tertiary ($\alpha_{13}$, $\delta_{13}$). 

\begin{figure}[H]
\centering
\includegraphics[width=7in]{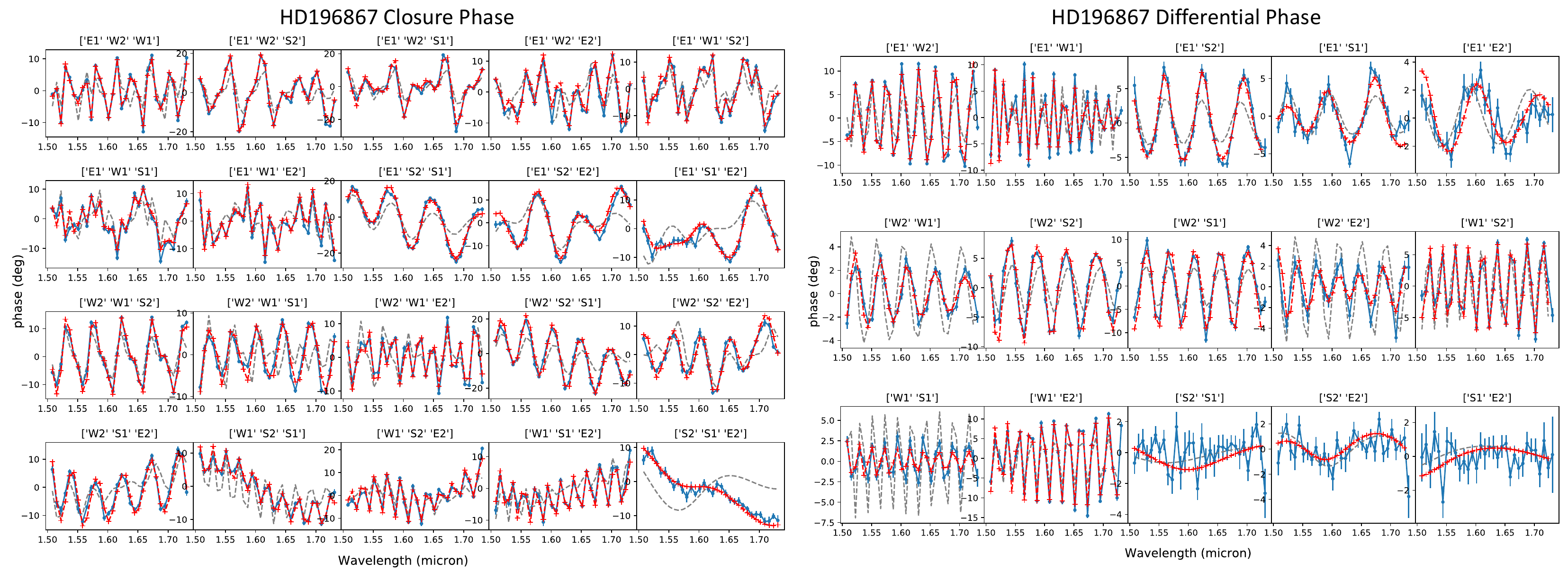}
\caption{We show an example fit of our triple model to our observables closure and differential phase of $\alpha$ Del from 2019Jul30. The two panels show our best fit (dashed red line with crosses) to closure phase (left) and differential phase (right). The dashed grey line shows the best-fit binary model, which is a significantly worse fit to our data than the triple model. Each square in the plots represents one of MIRC-X's 20 closing triangles or 15 baselines (the 6 telescopes are designated as E1-E2-W1-W2-S1-S2). This is a single 60-second measurement, of which we typically have $\sim$10 per observation of a target. }
\label{alphdel_example}
\end{figure}

\begin{figure}[H]
\centering
\includegraphics[width=6in]{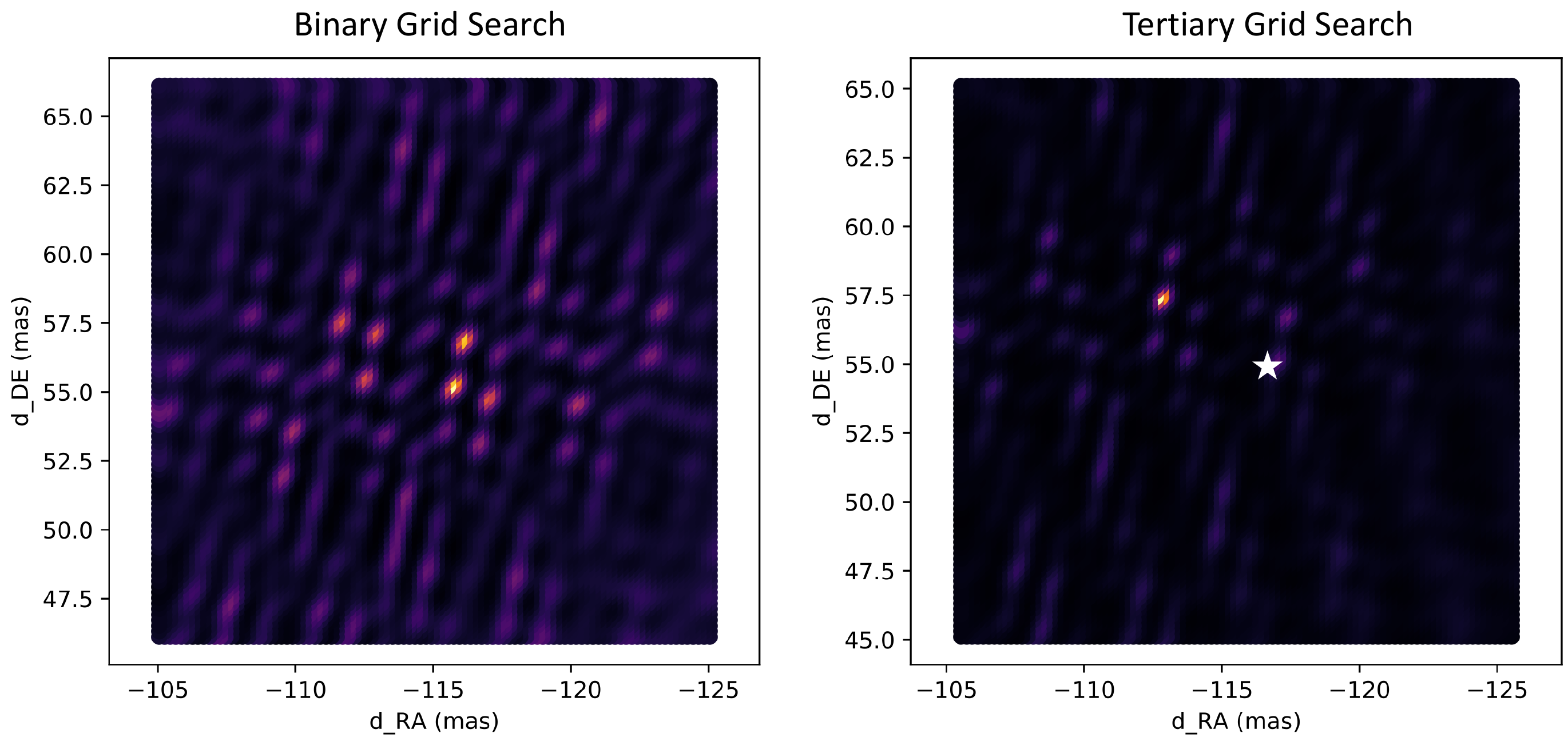}
\caption{We show example chi-square maps for the 2019Jul30 epoch of $\alpha$ Del. To search for flux from triple companions, we first perform a grid search for the brighter binary (left). Due to the signal from a tertiary companion, this initial $\chi^2$ map is quite messy. We then do a grid search for a third component, with a fit for the binary companion performed at each iteration (right). The location of the binary component is denoted by the star in the plot. Including this third component vastly improves the chi-square maps. Once we narrow in on the best solutions for each component on the grid, we perform a joint least-squares fit to refine the solutions. }
\label{alpdel_search}
\end{figure}

To find our best differential astrometry solution on a given night, we first perform a wide grid search in R.A. and declination with step sizes of 0.1 milli-arcseconds to find the minimum $\chi^2$ solution. We then perform a non-linear least squares fit using the $lmfit$ package in Python to narrow in on the best solution \citep{newville2016}. To search for a third component, we perform an additional grid search of the tertiary component and fit for the expected binary (initial guesses informed by our first grid search) on each point of this grid (Figure~\ref{alpdel_search}). After obtaining the approximate location of all three components, we perform a global fit with our triple complex visibility model starting from the best guesses from the grid searches. Figure \ref{alphdel_example} shows an example fit for one of our epochs of $\alpha$ Del.

We convert our astrometry solutions from differential R.A. and DEC to a separation and position angle east of north ($\rho$, $\theta$). After finding the best-fit $\rho$ and $\theta$, we estimate errors by performing a 2D scan in R.A. and declination around the best solution. The error ellipses we report are then a fit to the region bound by $\chi^2 = \chi^2_{best} \pm 1$. Since these error ellipse sizes are conservative estimates, we further scale these errors down to raise $\chi^2_{red}=1$ after performing our full set of orbital fits described in section \ref{sec:orbitfitting}.


\subsection{Wavelength Calibration}
\label{sec:observations-wavelength}
In most previous astrometric experiments (e.g., PTI-PHASES: \citealt{colavita1999,Muterspaugh2010}, Keck-ASTRA: \citealt{woillez2010}, VLTI-PRIMA: \citealt{sahlmann2013}, VLTI-GRAVITY: \citealt{gravity2017}), a differential delay line with laser metrology is used to measure the pathlength difference between broadband fringe packets from the two stars under study.  For ARMADA, we use a spectrometer with spectral resolution of $R=\frac{\lambda}{\Delta\lambda}\sim200$ so that each star's interferogram is about 200 fringes long. This allows the packets from the stars to overlap each other, creating distinctive wavelength dependent variations in the fringe phase. These variations can be directly related to the separation of the two stars and do not require any kind of differential delay line.  The downside is that we must know the effective wavelength of each spectral channel to high precision. MIRC-X is limited to a wavelength precision knowledge of $\frac{\Delta\lambda}{\lambda}\sim10^{-3}$ \citep{monnier2012}.  For a 100 mas binary, this would impose a limitation up to 100 $\mu$arcsec astrometry, too large for our ultimate goals of detecting the astrometric signal from orbiting exoplanets.

We employ an extra calibration step to bring our wavelength precision knowledge to the $10^{-4}$ level. Our team has utilized a custom-built 6-beam optical etalon system to calibrate our astrometry each ARMADA night. Each etalon consists of a thin, $\sim$2mm-thick piece of glass with parallel sides and 50\% reflective coatings on both sides.  By having an etalon in each beam being a slightly different thickness ($\Delta x = 6\pm2\mu$m),  our combiner detects multiple interference packets that look similar to a  binary star.  We measure the spectrometer wavelength using the same data pipeline and methodology as we are using for the science targets. We designed and built a special, thermally-stable etalon holder which is placed into the 6 CHARA beams each night. Our model for the etalon signal is similar to that of a binary signal in Equation~\ref{vis_eqn}, though the binary separation is now replaced by the differential thicknesses of the etalons:

\begin{equation}
    V = f_1 V_1 + \sum_{j=1}^{\infty} \Gamma f_{j+1} V_1 e^{-\pi i n_{\lambda} 2 j \Delta s / \lambda} .
\end{equation}

\begin{figure}[H]
\centering
\includegraphics[width=6in]{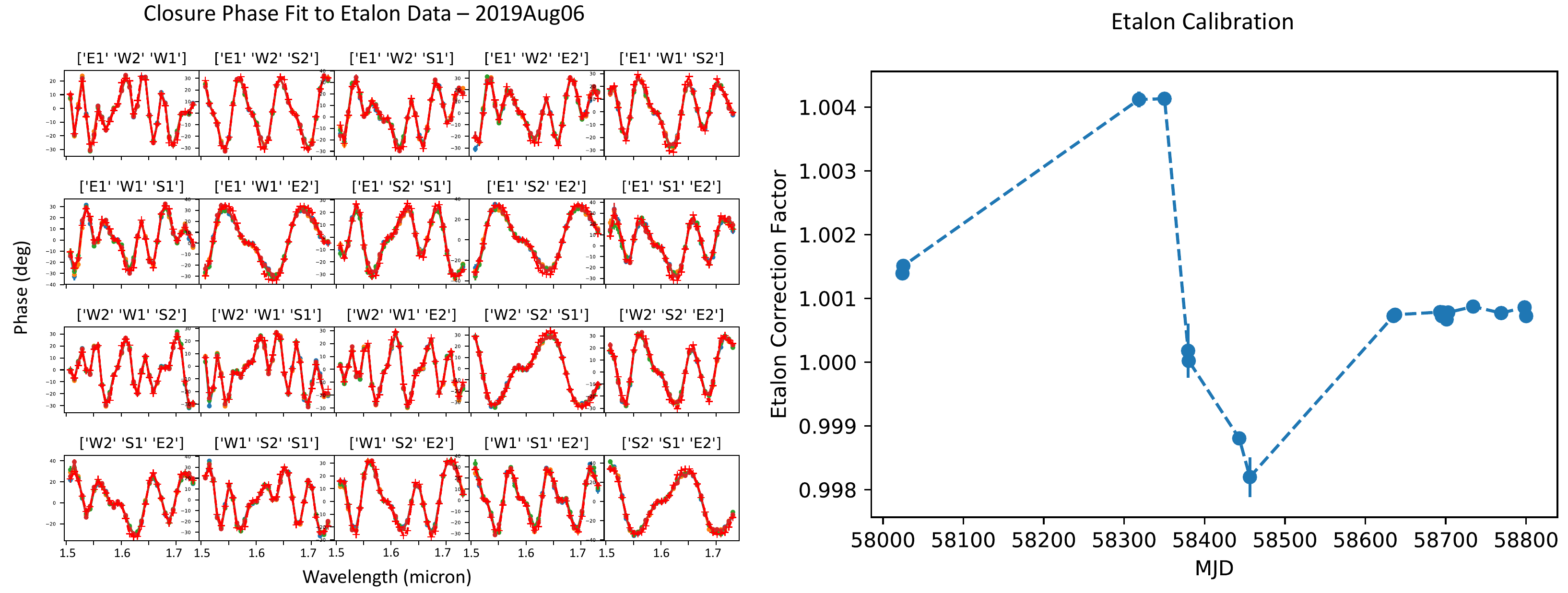}
\caption{(Left) We show an example fit from 2019Aug06 of our etalon data for wavelength calibration. The panels show a measurement of closure phase for each of MIRC-X's 20 closing triangles. It is nearly impossible to distinguish the etalon model (dashed red) from the data (blue circles). We take this etalon data each night, and monitor our wavelength solution against this reference. (Right) We show the correction factor values applied to the binary separation on each of our nights.}
\label{etalon_fit}
\end{figure}

In this equation $n_{\lambda}$ is the index of refraction for infrasil 301: $n_{\lambda} = [0.6961663 \lambda^2 / (\lambda^2-0.0684043^2) + 0.4079426 \lambda^2/(\lambda^2-0.1162414^2)+0.8974794 \lambda^2/(\lambda^2 - 9.896161^2) +1]^{1/2}$, $\Delta s$ is the differential thickness between two etalons of different beams, and $f_j$ is the fraction of flux in a given reflection. $V_1$ is the primary beam visibility, while now the bandwidth smearing parameter is defined as $\Gamma = \sinc[b(n_{\lambda} \Delta s / \lambda)]$. The etalons create an infinite number of reflections which can be modeled in theory, though we find that including more than two terms (j=1 and j=2) is not necessary for improving our $\chi^2$ fits. Figure~\ref{etalon_fit} shows an example dataset for a measurement of etalon signal from the internal STS source. 

A systematic error in the wavelength solution for a given night will affect the measured value of binary separation. We thus use our etalon calibration data to bring each night to the same astrometric scale. Each ARMADA night, we take etalon data to generate an astrometric correction factor to apply to the measured binary separations. Before the implementation of the internal STS source, these data were taken on-sky with the calibrator star. We fit for differential thickness between the 6 etalons for each beam (free parameters are five $\Delta s$ values between one reference beam, 2nd order polynomial for differential phase dispersion on each baseline, and the monochromatic flux ratio lost in a reflection). We then choose a high SNR reference night for ARMADA, and scale each night's binary separations based off this reference. The choice of the ARMADA reference night is arbitrary, since we are dealing only with internal astrometric consistency between ARMADA nights. In future work, we plan to use a shared binary source between MIRC-X and VLTI-GRAVITY in order to carry out an absolute calibration of MIRC-X wavelengths (GRAVITY achieves precise absolute wavelength calibration with an internal fourier transform spectrometer source). Currently, there is an estimated 0.25\% wavelength precision for MIRC-X \citep{monnier2012}. This systematic applies to the measured binary separations and needs to be taken into account when combining our presented astrometry with high precision astrometric data from other instruments. 

We calculate a single ``etalon correction factor" for each night based off of the slope of the 15 REFERENCE vs NIGHT etalon optical path differences (OPDs - one per baseline). Since our etalons are in a thermally-stable holder, any change in our etalon model comes from the MIRC-X wavelength solution. This scaling is then applied to the separation $\rho$ of the binaries for the night, and hence is done as the final step after finding the best differential astrometry solution described in the previous section. The separations that we report have our etalon calibration applied. In Fig \ref{etalon_fit} we show how this scale factor changes with time across our ARMADA datasets. The maximum effect that this correction has on our separations is a 4e-3 factor (400 micro-arcseconds for a 100mas binary), with the median scaling being at a 7.8e-4 factor (78 micro-arcseconds for a 100mas binary). The nights which have a correction of $\sim$4e-3 are worse than the expected systematic for MIRC from \citet{monnier2012}. The likely culprit of this high correction factor is a change in the detector readout windowing, which was set in 2018Feb to avoid a bad pixel in the fringe window. Since the spectrograph has a slightly different optical magnification across the field of view, this change of position of the fringe window in the detector would affect the fringe spatial frequency and thus the wavelength solution. This ``new" fringe window was unchanged for ARMADA data taken between 2018Feb - 2018Aug, which is consistent with the higher etalon correction values. The error bars on the etalon correction factor are computed by bootstrapping the etalon datasets in time, and taking the standard deviations of the resulting scale factors computed. The error bars on our etalon factors are all smaller than the reported errors for the separation from astrometry, and hence do not add error to the astrometric solution. 

\begin{figure}[H]
\centering
\includegraphics[width=6in]{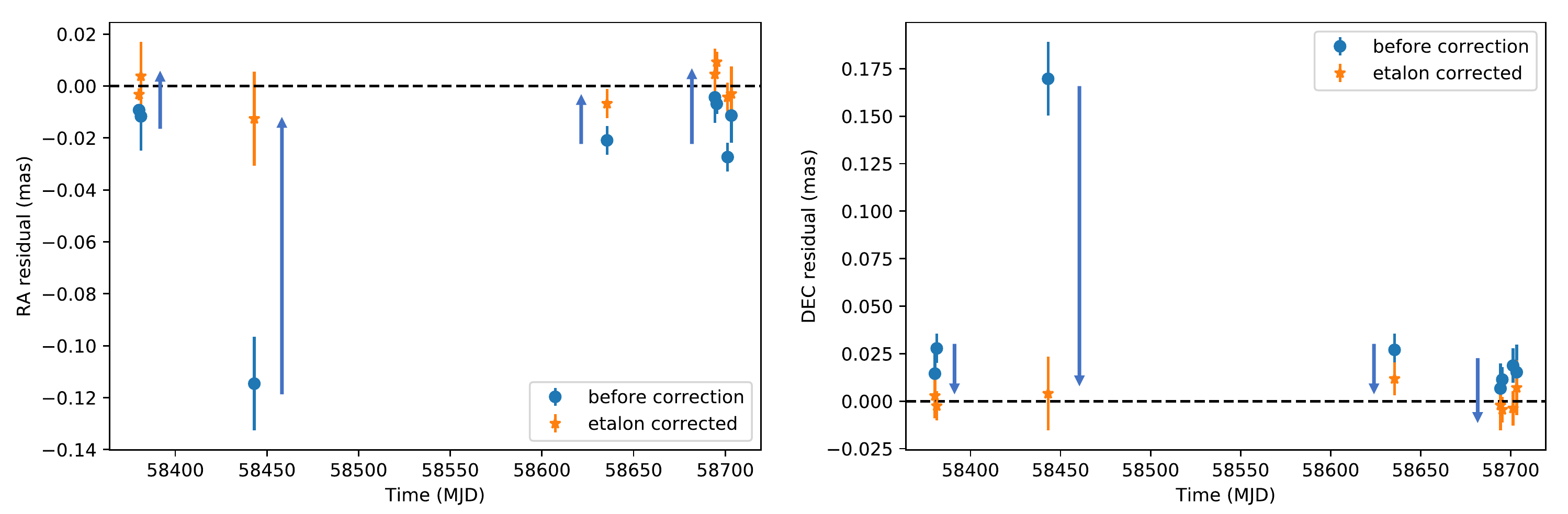}
\caption{We show the R.A. and DEC residuals of $\kappa$ Peg both before and after applying our etalon correction factor. The mean and median residual values both decrease, bringing us to our desired $\sim$10 micro-arcsecond precision which is needed to eventually detect substellar companions. The most significant correction in this case is a single outlier point which was $>$100 micro-arcseconds off from the expected orbit before the etalon factor is applied.}
\label{kappeg:correction}
\end{figure}

To test that our etalon wavelength correction scheme is working, we fit binary orbits for ARMADA both with and without the correction applied. Since our data only record a fraction of the outer orbital period, we also include historical data from the Washington Double Star (WDS) Catalog \citep{mason2001}. Table \ref{table:etalon} shows how the mean and median residuals to these orbit fits change (the full orbit fitting routine is presented in section \ref{sec:orbitfitting}). The median residual to the best fit orbit decreases significantly for all three targets after applying our etalon calibration. Though the median residual of $\nu$ Gem decreases significantly, it shows the least improvement in mean residual when applying the etalon correction (due to a few higher residual points). This added noise is possibly due to time-varying resolved structure in the Be-disk of this system, as discussed in Section \ref{sec:nugem}. Figure \ref{kappeg:correction} shows the residual fit in R.A. and DEC for $\kappa$ Peg, before and after the etalon correction is applied. 

\begin{table}[H]
\centering
\caption{Results of Etalon Wavelength Calibration}
\label{table:etalon}
\begin{tabular}{lcc}
\hline
\colhead{Target} & \colhead{No Etalon Correction} & \colhead{Etalon Correction Applied}   \\
\colhead{} & \colhead{mean / median residual} & \colhead{mean / median residual}   \\
\hline
$\kappa$ Peg & 45.0 / 24.9 $\mu$-as & 8.0 / 6.3 $\mu$-as \\
$\alpha$ Del & 61.4 / 36.3 $\mu$-as & 23.3 / 21.8 $\mu$-as \\
$\nu$ Gem & 48.5 / 35.9 $\mu$-as & 33.8 / 14.8 $\mu$-as \\
\hline
\end{tabular}
\end{table}

\section{Orbit Fitting}
\label{sec:orbitfitting}
Once we have our measured binary separations and position angles for each night, we are able to fit a Keplerian orbit to the data. Since we are aiming for high precision differential astrometry, we need to account for the precession of North when combining position angles measured by MIRC-X to historical data in the WDS catalog. The MIRC-X pipeline already defines the uv-plane in ICRS using the Python ``astropy.coordinates" package \citep{astropy2013,astropy2018}, thus accounting for Earth orientation. Note that we also make extensive use of the Numpy package in our Python routines \citep{harris2020array}. We correct the position angles of the WDS data to a common J2000 reference with 
\begin{equation}
    \rm{P.A(J2000)} = \rm{P.A.}(t) - 0.00557^{\circ} (t-2000) \sin{\alpha} / \cos{\delta}, 
\end{equation}
where $t$ is the year of observation \citep{siregar2010}. 

The Campbell elements ($\omega$, $\Omega$, $e$, $i$, $a$, $T$, $P$) describe the Keplerian motion of one star of a binary system relative to the other. Those symbols have their usual meanings where $\omega$ is the longitude of the periastron, $\Omega$ is the position angle of the ascending node, $e$ is the eccentricity, $i$ is the orbital inclination, $a$ is angular separation, $T$ is a time of periastron passage, and $P$ is orbital period. For near-circular orbits $T$ and $\omega$ become ill-defined, adding seemingly large errors to these orbital parameters as the two parameters are correlated. In these cases we also report a time of maximum radial velocity $T_{rv,max}$, which is more tightly constrained. When including RV data, we also fit to the semi-amplitudes $K$ and system velocity $\gamma$. The longitude of periastron $\omega$ is traditionally reported for the secondary when fitting to visual binary orbits alone. The convention when combining RV orbits is to report $\omega$ of the primary, which is flipped by 180$^{\circ}$. We have RV data for both $\kappa$ Peg and $\alpha$ Del, and hence we report the $\omega$ of the primary for these orbits. For $\nu$ Gem we only present the visual orbit, and so we report $\omega$ of the secondary (noted in the table of orbital elements). For visual orbits, there is a 180 degree ambiguity between $\omega$ and $\Omega$. Our RV information breaks this degeneracy for $\kappa$ Peg and $\alpha$ Del, and we report the $\Omega < 180^{\circ}$ for $\nu$ Gem. 

For nonlinear least-squares fitting, we use the Thiele-Innes elements to describe our Keplerian orbits. As described in \citet{Wright2009}, these elements convert ($\omega$, $\Omega$, $i$, $a$) to linear parameters (A, B, F, G). When fitting a system of three components, we assume the 3-body system is hierarchical with the wide companion orbiting the center-of-mass of the inner pair. This means that our orbit model is simply a sum of the outer + inner Keplerian orbits. Since the outer orbits we present in this paper are significantly larger than the inner orbits ($>$200 times larger in orbital period), this hierarchical model is a reasonable assumption. When flux from the third component is not detected, our orbital elements are then describing the ``wobble" motion of one star about the center-of-mass of the inner orbit. In this case the angular semi-major axis $a_{wob}$ of the tertiary component describes the size of the wobble motion, where one would need to know the mass ratio to figure out the true angular semi-major axis of the inner pair. 
In the case where we detect flux from all three components, we then modify our orbital elements to also include the mass ratio since we are then able to measure $a_1$ and $a_2$ of the inner semi-major axis ($a_{inner} = a_1 + a_2$). We then do a joint fit to the outer orbit, the wobble motion, and the inner visual orbit. 

\begin{figure}[H]
\centering
\includegraphics[width=6in]{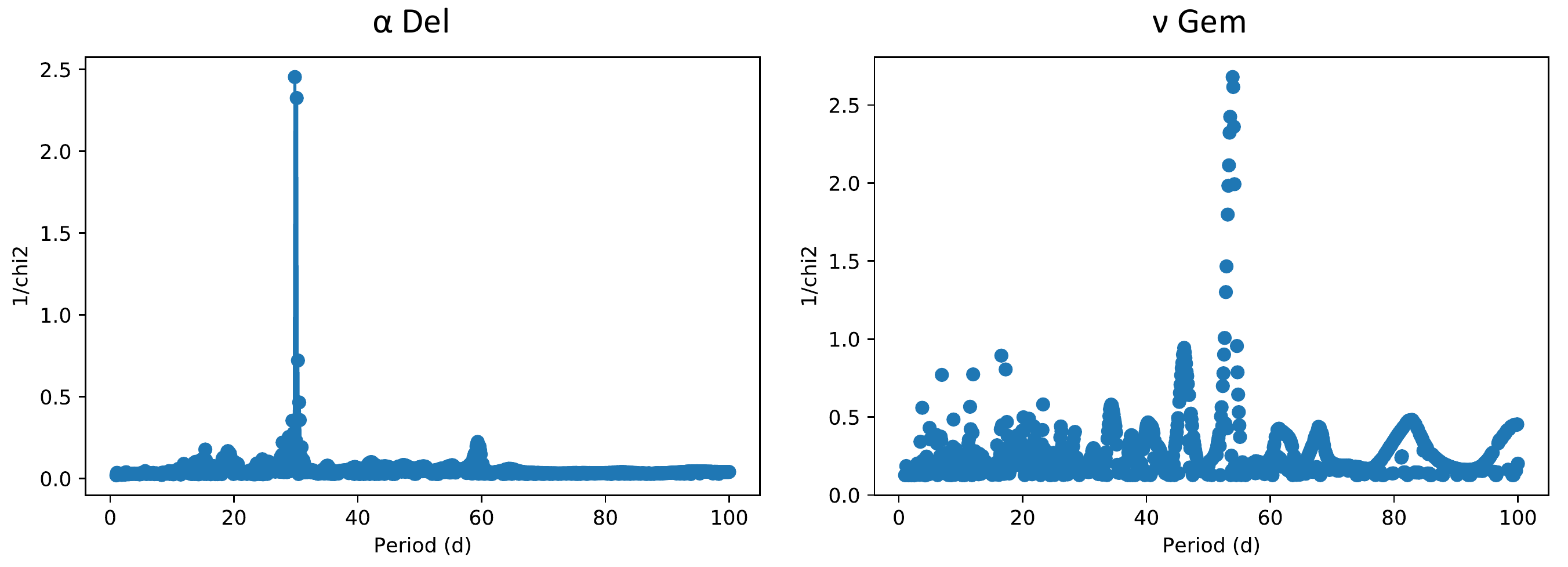}
\caption{We show our period search results for the new detections of inner tertiary companions to binaries $\alpha$ Del and $\nu$ Gem. Since the period and eccentricity are nonlinear parameters in our orbit fits, we search for extra components by varying the period on a grid and searching for circular tertiary orbits. For $\alpha$ Del we see a clear signal at a period of about 30 days, while $\nu$ Gem shows a strong detection at a period of about 54 days. The reduced $\chi^2$ values are $<1$ in these cases, as this search occurs before scaling the final ARMADA error bars. }
\label{periodogram}
\end{figure}

We again use the Python $lmfit$ package for non-linear least-squares fitting of our data \citep{newville2016}. To constrain the outer binary orbits, we include historical data from WDS. Based on the large scatter of these data about their best fits, we assign circular errors of radius 10 milli-arcseconds to the WDS data. We use the ORB6 catalog for initial guesses of our orbital parameters for the outer pair. Once we find the best fit for the outer binary, we begin searching for the inner companion. To do so, we vary the inner orbital period and fit circular orbits to each fixed period as shown in Figure \ref{periodogram}. Once the best inner period is detected, we refine our search further by performing a joint outer + inner fit with all orbital parameters varying.

Error bars for the fitted orbital parameters are normally estimated in $lmfit$ from the covariance matrix, but since the orbital elements $P$ and $e$ are nonlinear we instead determine posterior distributions on our orbital parameters with a Markov chain Monte Carlo (MCMC) fitting routine. We carry out MCMC fitting using the Python package \textit{emcee} developed by \citet{Foreman2013}. We use our best-fit orbital elements as a starting point for our 2*N$_{\rm{params}}$ walkers, where the starting point for each walker is perturbed about its best fit value. We assume uniform priors on all of our orbital elements. The quoted error bars on our orbital elements are the standard deviations of the posterior distributions, and corner plots of the posteriors for the inner and outer orbits show correlations between parameters. 

\section{Kappa Peg: Verifying ARMADA Astrometry}
\label{sec:kappeg}
Before presenting our new astrometric detections in systems $\alpha$ Del and $\nu$ Gem, we wanted to test our ARMADA calibration scheme on a well-known system to verify our precision astrometry. \citet{muterspaugh2006} published a high precision orbit on the triple star system $\kappa$ Peg (HD 206901, HIP 107354, WDSJ21446+2539) with their interferometric PHASES survey. This system consists of a wide $\sim$0.2" pair of F5 subgiants, and an inner component around the brighter star with a 6-day orbital period. We observed this tertiary system over the course of a year with our ARMADA survey, and compare our inner orbit to that obtained by \citet{muterspaugh2006}. For historical reasons, the brighter component in H-band was designated the B component of the outer A-B system -- a nomenclature that the previous work kept. To stay consistent with most previous work on $\kappa$ Peg and make a direct comparison of orbital elements, we will keep this designation when reporting orbital elements. Hence the brighter star is the Ba+Bb inner subsystem.

\begin{figure}[H]
\centering
\includegraphics[width=4.5in]{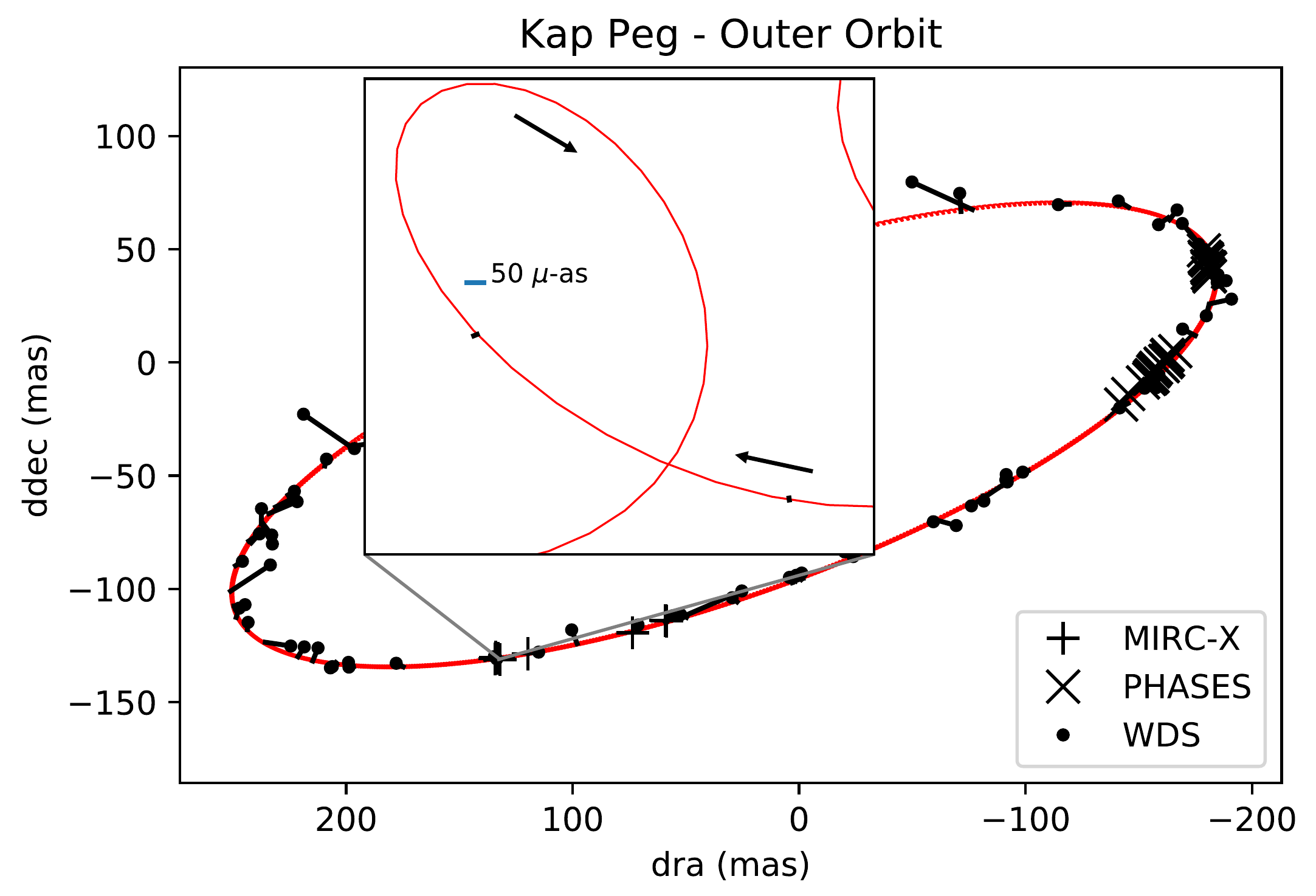}
\caption{$\kappa$ Peg is a triple star followed by both ARMADA and PHASES surveys. We show the data from MIRC-X and PHASES together, along with the best fit to the outer orbit. We also plot historical data obtained from the WDS catalog. The zoomed inset to a portion of our ARMADA data gives an idea of the size of our error ellipses. }
\label{kappeg_full}
\end{figure}

\begin{figure}[H]
\centering
\includegraphics[width=6in]{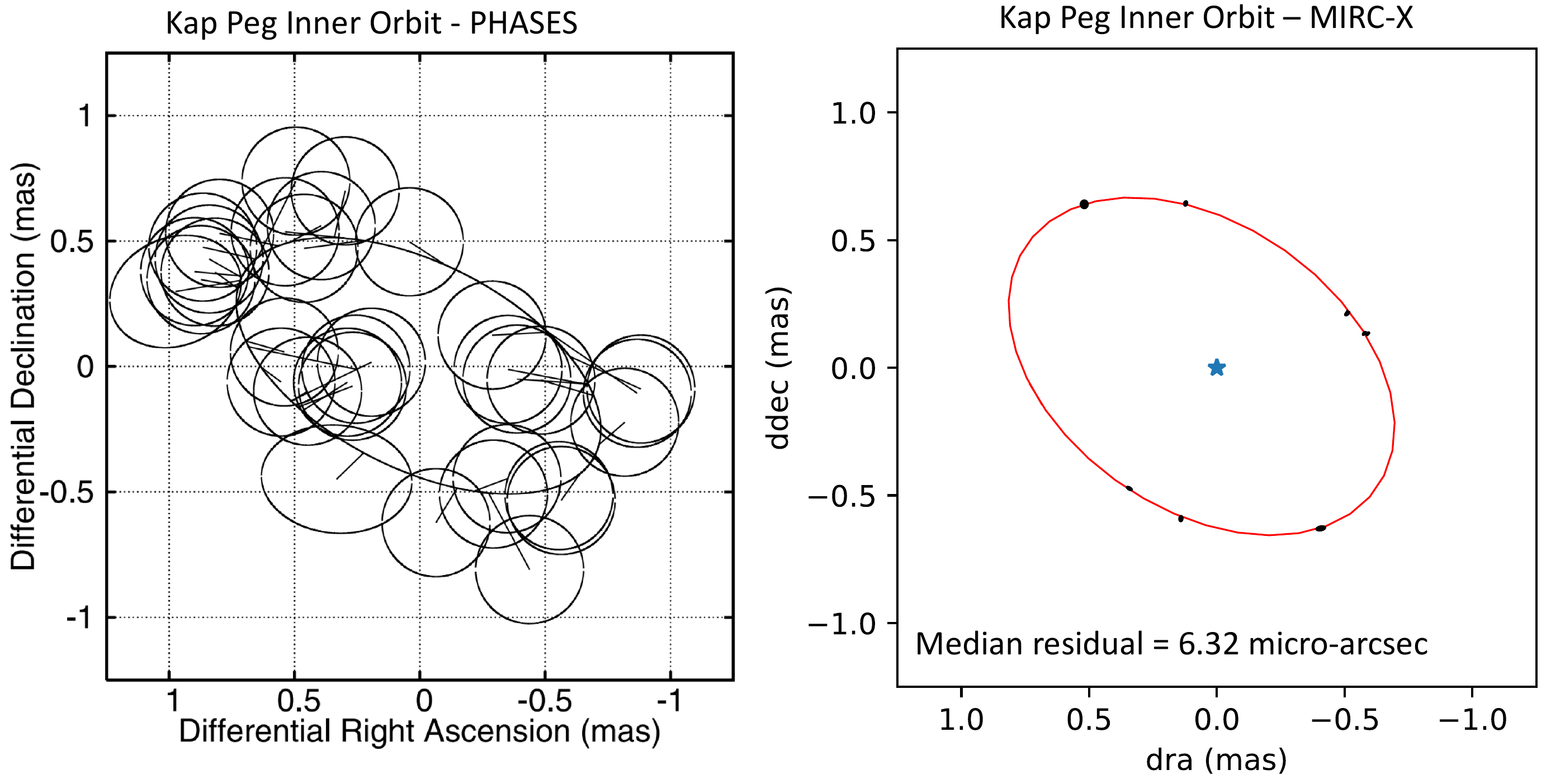}
\caption{We show our best-fit orbit to the inner "wobble" of $\kappa$ Peg due to the tertiary companion, after subtracting out the outer binary motion. We compare the results of the PHASES best-fit orbit from \citet{muterspaugh2006} (left) to our MIRC-X+WDS fit (right). Over the $\sim$year time baseline of following this object with MIRC-X we are achieving a median residual level of 6.32 micro-arcseconds, which is very promising for our ultimate goal of detecting giant planets in binary systems with the ARMADA survey. }
\label{kappeg_inner_comparison}
\end{figure}

We obtained 8 new data points for this system throughout 2018-2019, taken after the MIRC-X optics upgrade of 2018Sep. In Figure~\ref{kappeg_full}, we show the MIRC-X and PHASES data on our best fit to the outer AB orbit of $\sim$11.5 years. We first ignore the PHASES data for a comparison, fitting only to the MIRC-X data and historical WDS data -- important for constraining the outer orbit. Our high precision astrometry captures the motion of the brighter component about the center-of-mass of its inner pair. We show our fit to this ``wobble" motion after subtracting out the binary motion, in order to visually compare the astrometry between the PHASES and MIRC-X datasets. We present our best-fit MIRC-X orbital elements in Table \ref{kappeg:orbitelements}, along with the previously published elements for comparison. Figure~\ref{kappeg_inner_comparison} is a good demonstration of the improvement in interferometric data quality over the past decade between PHASES and ARMADA. Our median residual to the binary+wobble motion is an amazing 6.32 micro-arcseconds after we apply our etalon wavelength calibration, when fitting to ARMADA+WDS datasets. This demonstrated precision is about a factor of 10 improvement over PHASES, which is very promising for our ultimate goal with the ARMADA survey to detect circumstellar giant planets in A/B-type binary systems.

We also combine our new astrometric datasets with the PHASES data and radial velocity data presented in \citet{muterspaugh2006} for a full combined fit. Our update to the best orbit of $\kappa$ Peg is presented in the final column of Table \ref{kappeg:orbitelements}. Note that \citet{muterspaugh2008} reported updated orbital elements for this system, as there was a sign flip present in the original analysis. We use these updated orbital elements for the PHASES+RV column of this table. Since we view the ``wobble" motion of the bright component, along with the single-line RV motion of this component, we are able to compute orbital parallax for the system. Our value agrees well with the Hipparcos distance of $34.2 \pm 0.9$ pc \citep{vanLeeuwen2007}, as well as the previously published value for PHASES. We also update the masses of all three components in the system. Figure~\ref{kappeg_corner} shows the corner plots of our posterior distributions from the MCMC routine. We see no significant correlations between the parameters of the inner orbit with the parameters of the outer orbit -- hence we split these corner plots into outer and inner orbital elements to increase clarity. In general, the outer long-period orbit shows greater correlations between the orbital parameters. This is not surprising, as the coverage is not as complete for the $\sim$11.5 year period as compared to the inner 6-day period. For near circular orbits, the correlation between the angle of periastron and the time of passage through periastron is expected -- as these parameters become ill-defined. This is especially apparent in the inner orbit. There is also a strong correlation between the inclination and semi-major axis for the outer orbit, which is normal for partial coverage of a visual orbit.

We were not able to detect the flux from the Bb component of the system. Our measured mass for this component of $0.814\pm0.046 M_{\odot}$ could be consistent with that of a white dwarf remnant of a massive star, though \citet{muterspaugh2006} claim to see evidence of a third set of lines in their Keck-HIRES spectra (although RVs of this third set were not measured in that work). This makes it more likely that the Bb component is a late G or early K dwarf, which should imply a flux ratio of $\sim$7\% with respect to the Ba component in the near infrared. MIRC-X is easily capable of detecting companions of this flux ratio, though the signal from the near 1:1 wide binary dominates in this case. Our rough visibility calibration scheme for the ARMADA survey also increases the difficulty of detecting faint nearby companions (expected semi-major axis is $\sim$2.5 mas). Though we found some tentative detections of the third companion in a few of our nights, they all appeared to be related to residual structure in the $\chi^2$ maps. These tentative detections also did not align with the expected position angle from the ``wobble" orbit, hence we concluded that they were not real.

\begin{figure}[H]
\centering
\includegraphics[width=6in]{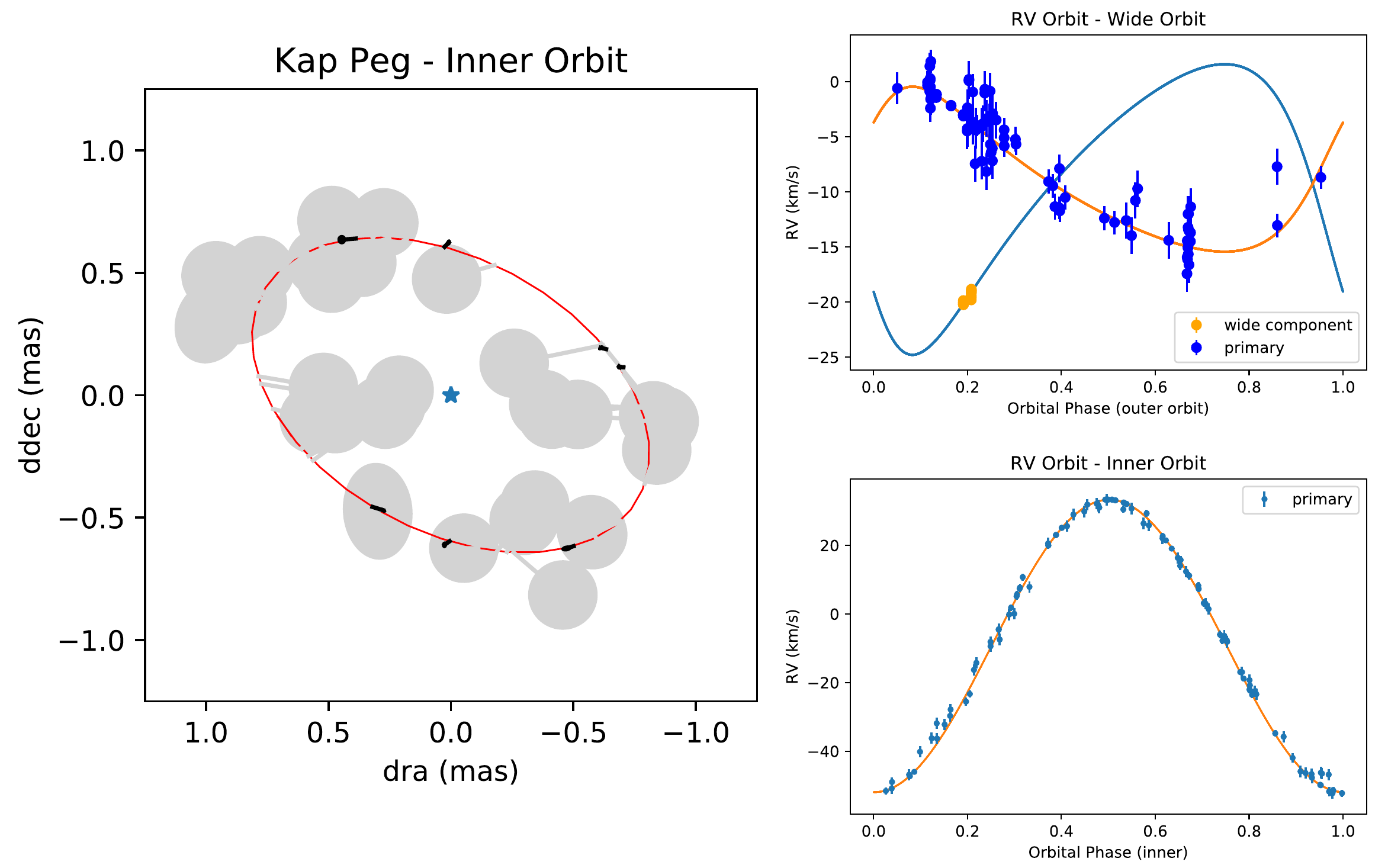}
\caption{For a final orbital fit, we combine our new MIRC-X data with the historical data and PHASES/RV data presented in \citet{muterspaugh2006}. (Left) We plot the bright component going around the center-of-mass of the inner pair, once the outer binary motion is subtracted out. The light grey ellipses are data from PHASES, while the smaller black ellipses are the new data from MIRC-X. We show RV data of the primary and secondary phased to the outer orbital period of 11.6 years (upper right). Here the RV motion of the bright primary due to the inner companion is subtracted out. We also show data of the brighter component phased to the 5.9 day inner period (lower right).}
\label{kappeg_inner}
\end{figure}

\begin{figure}[H]
\centering
\includegraphics[width=7in]{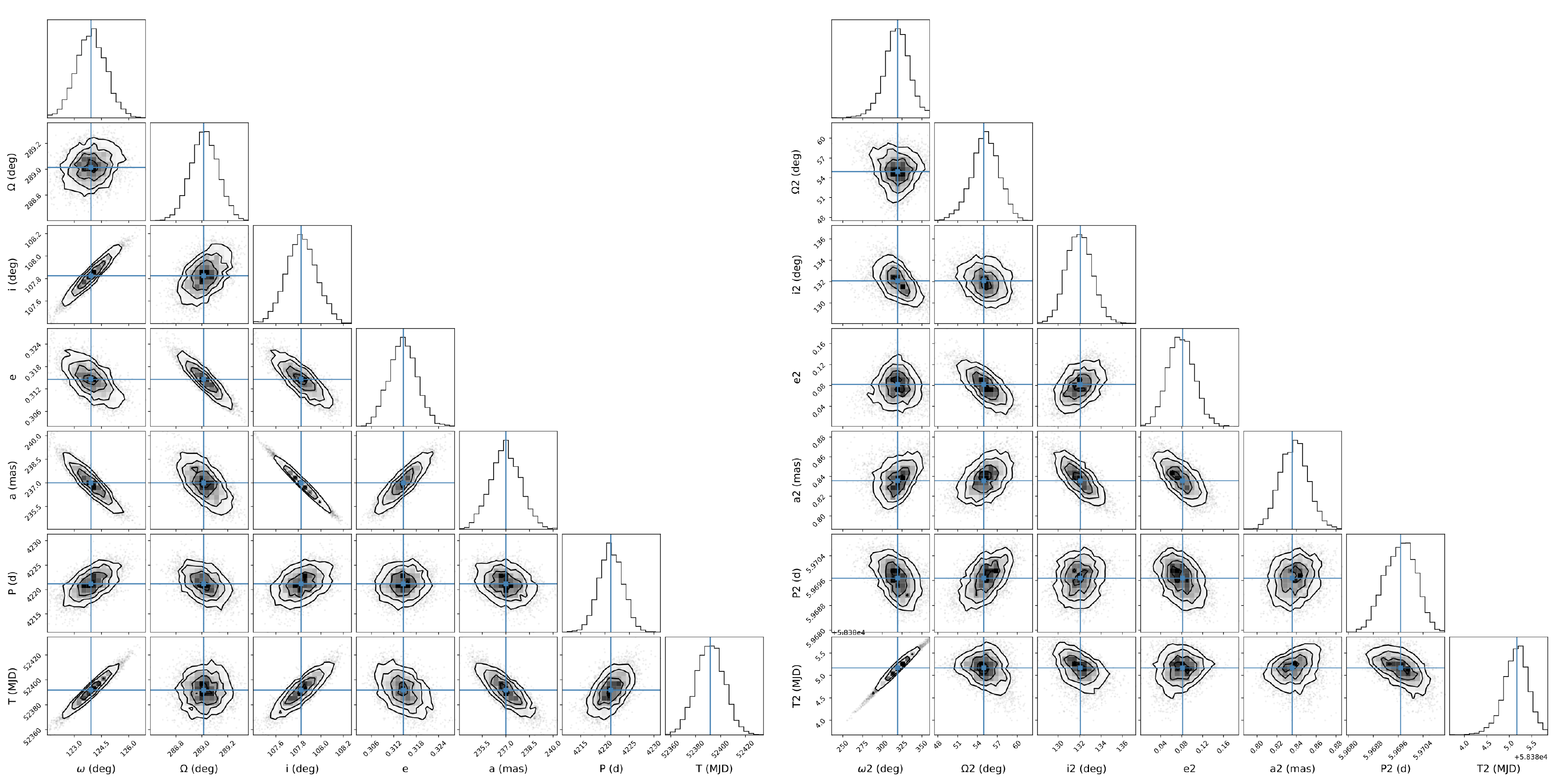}
\caption{We show corner plots of the MCMC posterior distributions produced from \textit{emcee} for the outer (left) and inner (right) orbital elements for $\kappa$ Peg. Since the outer long-period orbit is less well-constrained than the short-period inner orbit, there are more obvious correlations between the parameters. Since the inner orbit is near circular, there is a strong correlation between $\omega_2$ and $T_2$. }
\label{kappeg_corner}
\end{figure}

\begin{table}[H]
\centering
\caption{$\kappa$ Peg Astrometry\tablenotemark{a}}
\label{table:astrometry-kappeg}
\begin{tabular}{lcccccc}
\hline
\colhead{UT Date} & \colhead{MJD\tablenotemark{b}} & \colhead{sep (mas)}   & \colhead{P.A. ($^\circ$)} & \colhead{error major axis (mas)} & \colhead{error minor axis (mas)} & \colhead{error ellipse P.A. ($^\circ$)} \\
\hline
2018Sep19 & 58380.2384 & 128.295 & 152.633 & 0.034 & 0.019 & 280.93 \\
2018Sep20 & 58381.2551 & 128.271 & 152.877 & 0.024 & 0.014 & 59.70 \\
2018Nov21 & 58443.1002 & 140.177 & 148.379 & 0.030 & 0.028 & 1.34 \\
2019Jun01 & 58635.5138 & 175.684 & 137.070 & 0.022 & 0.014 & 309.41 \\
2019Jul30 & 58694.3468 & 186.101 & 134.800 & 0.021 & 0.016 & 346.87 \\
2019Jul31 & 58695.3394 & 186.426 & 134.495 & 0.017 & 0.012 & 300.65 \\
2019Aug06 & 58701.3077 & 187.472 & 134.261 & 0.019 & 0.012 & 318.42 \\
2019Aug08 & 58703.3354 & 186.810 & 134.175 & 0.023 & 0.017 & 354.75 \\
\hline
\end{tabular}
\tablenotetext{a}{Note that we are reporting the position of the fainter companion relative to the brighter star. In keeping the designation of most previous work on $\kappa$ Peg, this is actually the B component relative to the A component, since these stars were flipped. }
\tablenotetext{b}{Modified Julian Date (MJD) = Julian Date (JD) - 2400000.5}
\end{table}

\begin{table}[H]
\centering
\caption{$\kappa$ Peg: Best fit orbital elements}
\label{kappeg:orbitelements}
\begin{tabular}{lccc}
\hline
\colhead{} & \colhead{MIRC-X Visual Orbit} & \colhead{PHASES+RV} & \colhead{Combined}\\
\hline
$P$ (d) & $4221.18 \pm 1.43$ & $4224.76 \pm 0.74$ & $4222.420 \pm 0.43$ \\
$T$ (MJD) & $52391.76 \pm 6.42$ & $52401.52 \pm 0.96$ & $52398.17 \pm 1.17$ \\
$e$ & $0.3146 \pm 0.0021$ & $0.3140 \pm 0.0011$ & $0.31306 \pm 4.8\cdot10^{-4}$ \\
$\omega$ $(^\circ)$ & $123.92 \pm 0.46$ & $124.666 \pm 0.064$ & $124.449 \pm 0.086$ \\
$\Omega$ $(^\circ)$ & $289.015 \pm 0.062$ & $289.037 \pm 0.021$ & $289.052 \pm 0.013$ \\
$i$ $(^\circ)$ & $107.825 \pm 0.076$ & $107.911 \pm 0.029$ & $107.914 \pm 0.015$ \\
$a$ (mas) & $236.99 \pm 0.54$\tablenotemark{a} & $235.22 \pm 2.3$ & $236.345 \pm 0.094$\tablenotemark{a} \\
\hline
$P_2$ (d) & $5.96967 \pm 2.2\cdot10^{-4}$ & $5.9714971 \pm 1.3\cdot10^{-6}$ & $5.9714973 \pm 1.4\cdot10^{-6}$ \\
$T_2$ (MJD) & $58385.17 \pm 0.14$ & $52402.22 \pm 0.10$ & $52402.24 \pm 0.15$ \\
$e_2$ & $0.082 \pm 0.015$ & $0.0073 \pm 0.0013$ & $0.00166 \pm 3.8\cdot10^{-4}$ \\
$\omega_2$ $(^\circ)$ & $179.33 \pm 8.39$ & $179.0 \pm 6.0$ & $179.39 \pm 11.14$ \\
$\Omega_2$ $(^\circ)$ & $234.93 \pm 1.16$ & $244.1 \pm 2.3$ & $239.81 \pm 0.52$ \\
$i_2$ $(^\circ)$ & $132.07 \pm 0.71$ & $125.7 \pm 5.1$ & $129.49 \pm 0.83$ \\
$a_2$ (mas) & $0.8357 \pm 0.0082$\tablenotemark{a} & $0.828 \pm 0.040$ & $0.873 \pm 0.010$\tablenotemark{a} \\
$T_{2, \rm{rv,max}}$ (MJD) & $58391.72 \pm 0.03$ & -- & $58391.665\pm0.001$ \\
\hline
$K_A$ (km~s$^{-1}$) & -- & $11.78 \pm 0.24$\tablenotemark{b} & $13.19 \pm 0.16$ \\
$K_B$ (km~s$^{-1}$) & -- & $7.37 \pm 0.20$\tablenotemark{b} & $7.48 \pm 0.12$ \\
$K_{B_a}$ (km~s$^{-1}$) & -- & $42.8 \pm 0.4$\tablenotemark{b} & $42.610 \pm 0.011$ \\
$\gamma$ (km~s$^{-1}$) & -- & $-9.41 \pm 0.25$ & $-9.27 \pm 0.11$ \\
\hline
\hline
Physical Properties \\
\hline
$f_A/f_B$ (H-band) & $1.332 \pm 0.004$ & -- & $1.332 \pm 0.004$ \\
$f_A/f_B$ (K-band) & -- & $1.1912 \pm 0.0011$\tablenotemark{c} & $1.1912 \pm 0.0011$\tablenotemark{c} \\
$d$ (pc) & -- & $34.57 \pm 0.21$ & $33.88 \pm 0.33$ \\
$M_A$ ($M_{\odot}$) & -- & $1.533 \pm 0.050$ & $1.391 \pm 0.044$ \\
$M_{B_a}$ ($M_{\odot}$) & -- & $1.646 \pm 0.074$ & $1.616 \pm 0.049 $ \\
$M_{B_b}$ ($M_{\odot}$) & -- & $0.825 \pm 0.059$ & $0.835 \pm 0.026$ \\
\hline
\end{tabular}
\tablenotetext{a}{$\pm0.25\%$ from absolute wavelength precision}
\tablenotetext{b}{RV semi-amplitudes are not reported in \citet{muterspaugh2008}. We compute these from the elements given in that paper.}
\tablenotetext{c}{Flux ratios in K are derived from Keck AO imaging in \citet{muterspaugh2006}, rather than PHASES data.}
\end{table}

\section{Two Newly Detected Compact Triples}
\subsection{$\alpha$ Del}
\label{sec:alpdel}
The star $\alpha$ Del (HD196867, HIP101958, WDSJ20396+1555) is bright (V = 3.8, H = 3.9) with a spectral type of B9IV \citep{morgan1973}. 
$\alpha$ Del was first determined to have a binary companion by \citet{wickes1975}. That study used the Mt. Wilson 60-inch telescope to find a $\sim$200 milli-arcsecond binary companion with a magnitude difference of about 2 in the visual passband. \citet{mcalister1977} then followed up on this new system with speckle interferometry on the Mayall 4-meter telescope, confirming the new detection and in subsequent studies adding additional points to the binary orbit. The binary orbit has since been well constrained with speckle data, with WDS reporting nearly 100 data points from 1974-2015. The Sixth Catalog of Orbits of Visual Binary Stars (ORB6, \citealt{malkov2012}) lists a 17 year period for the binary orbit, with an eccentricity of 0.47. The ORB6 orbit is graded a 2 on a scale of 1-5, meaning most of the revolution is covered and the orbit quality is considered good. 

\begin{figure}[H]
\centering
\includegraphics[width=7in]{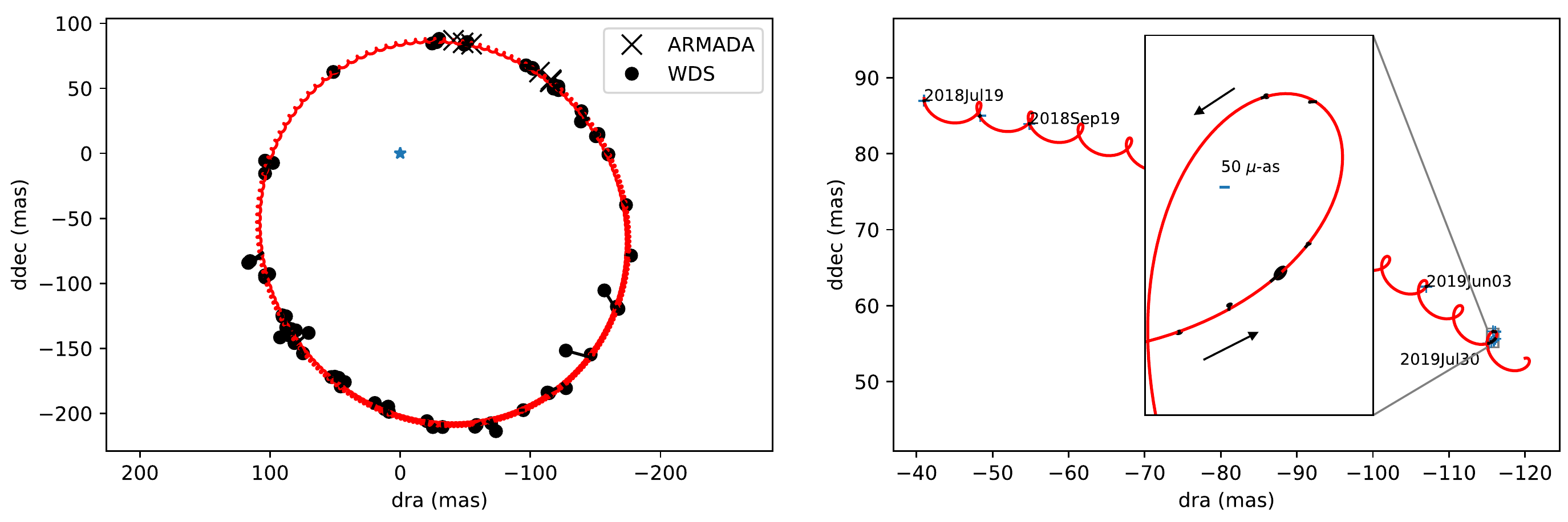}
\caption{(Left) We show the outer binary orbit of $\alpha$ Del as the Ba component moves relative to the A component fixed at the origin. Though the WDS data do not have enough precision to help constrain the inner orbit, we include these data in order to constrain the outer binary orbit. (Right) Showing only the new ARMADA data, clearly we detect an added astrometric wobble due to the presence of the Bb component.}
\label{hd196867_outer_orbit}
\end{figure}

We report the discovery of a third object in this system. We find that the B component is itself a short period subsystem, with an orbit of 30 days. This additional ``wobble" can clearly be seen with our astrometric precision of the outer binary orbit in Figure \ref{hd196867_outer_orbit}. We also detect the flux from the newly discovered Bb component in our MIRC-X data, which allows us to see the motion of both Ba and Bb components in the new pair about their center-of-mass. With 11 epochs, we are able to fit a full triple orbit model. To help constrain the outer orbit, we also include 70 epochs of historical data from WDS. 

To better constrain the new inner orbit and obtain a measurement of orbital parallax, we collected new RV data from the Tennessee State University 2~m Automated Spectroscopic Telescope (AST) and its
echelle spectrograph at the Fairborn Observatory \citep{ew07}.
See \citet{gardner2018} for a discussion of the spectroscopic observations from this telescope and their velocity reduction. From 2020 January through October we obtained 55 RVs, covering 
the orbit of the inner pair. We identified 2 sets of lines from the spectra - a broad-line component ($v$~sin~$i$ $\sim$145 km~s$^{-1}$) for the primary detected with a line list for A stars, and a second very narrow component ($v$~sin~$i$ $\sim$7 km~s$^{-1}$), which can be seen with both an A star line list and a solar star line list. The first RV component is stationary, and thus we attribute it to the primary A-component. The second RV component shows the $\sim$30-day motion of the Ba-Bb pair, which can be used to compute $a_1 \sin i$ of the inner orbit. It is not immediately obvious whether the measured RV is that of the Ba or Bb component. Since it is detected in both the A-star and solar line lists, this suggests a component which has a late A or early F spectral class. Our orbital parallax and mass determinations in Table \ref{physicalelements:alpdel} are consistent with this picture if the measured RVs are that of the less massive Bb component. The measured masses in this scenario would imply a mid-to-early A star for the Ba component. A rapidly rotating A-star would make this component difficult to detect in the composite spectra, and would explain why this third component is not easily visible despite it being the more massive star of the inner pair. A close inspection of the
mean composite line profile of the stars, which in total is only about 1 percent deep and was obtained with the A-star line list, shows a very weak, broad asymmetry, with a $v$~sin~$i$ of roughly 60 km~s$^{-1}$ that shifts in the opposite velocity direction to the component that we have identified as Bb. This provides supporting evidence for our conclusion that the narrow lines are from component Bb.

Since we know the inclination, the angular semi-major axis of Ba+Bb, and the motion of Ba about the center-of-mass from astrometry, and the $a_2 \sin i$ of Bb from RV, we are able to measure $a_2$ of Bb in both physical and angular units. This gives us a measurement of orbital parallax which can be used to compute masses of all three components without depending on an outside measurement of distance to the system. Our distance of $78.5\pm2.2$ pc is consistent with the Hipparcos value of $77.8 \pm 2.7$ pc \citep{vanLeeuwen2007}. The Gaia DR2 parallax of $66.8 \pm 2.4$ pc \citep{gaia2018} does not agree well with our measured orbit, though this is likely because DR2 does not yet take multiplicity into account when computing the parallax. We measure three masses of $3.83\pm0.33$ M$_{\odot}$, $1.82\pm0.15$ M$_{\odot}$, and $1.49\pm0.12$ M$_{\odot}$ for the primary, Ba component, and Bb component respectively. We report the orbital elements for the outer pair and this newly detected tertiary companion in Table \ref{orbitelements:alpdel}. Table \ref{physicalelements:alpdel} gives the physical properties of these stars which can be deduced from the single-line RV + visual orbits. Figure~\ref{alpdel_corner} shows posterior distributions of the outer and inner orbital elements. As for $\kappa$ Peg in the previous sections, there are no significant correlations between the parameters of the inner orbit with the parameters of the outer orbit. We again split the parameters up in the plot for clarity. The inner orbit of $\alpha$ Del is particularly well constrained. Since the inner orbit it near circular, there is the expected correlation between $T$ and $\omega$.

\begin{table}[H]
\centering
\caption{$\alpha$ Del A-Ba Astrometry}
\label{table:astrometry-hd196867}
\begin{tabular}{lcccccc}
\hline
\colhead{UT Date} & \colhead{MJD} & \colhead{sep (mas)}   & \colhead{P.A. ($^\circ$)} & \colhead{error major axis (mas)} & \colhead{error minor axis (mas)} & \colhead{error ellipse P.A. ($^\circ$)} \\
\hline
2018Jul19 & 58318.3213 & 96.087 & 334.836 & 0.042 & 0.020 & 315.00 \\
2018Aug21 & 58351.2246 & 97.689 & 330.437 & 0.052 & 0.027 & 314.05 \\
2018Sep19 & 58380.2076 & 100.226 & 326.906 & 0.055 & 0.031 & 277.72 \\
2019Jun03 & 58637.4281 & 123.853 & 300.394 & 0.020 & 0.014 & 315.03 \\
2019Jul29 & 58693.2785 & 127.582 & 295.618 & 0.027 & 0.018 & 325.10 \\
2019Jul30 & 58694.2693 & 127.981 & 295.627 & 0.025 & 0.014 & 330.07 \\
2019Jul31 & 58695.2677 & 128.360 & 295.648 & 0.056 & 0.042 & 302.26 \\
2019Aug01 & 58696.3156 & 128.612 & 295.689 & 0.020 & 0.016 & 315.12 \\
2019Aug06 & 58701.2788 & 129.023 & 296.060 & 0.015 & 0.011 & 313.05 \\
2019Aug08 & 58703.2806 & 128.781 & 296.132 & 0.022 & 0.021 & 34.96 \\
2019Nov12 & 58799.0774 & 134.547 & 289.244 & 0.050 & 0.032 & 42.23 \\
\hline
\end{tabular}
\end{table}

\begin{longtable}{lc}
\caption{$\alpha$ Del Radial velocities of Bb component}
\label{table:rv-hd196867}\\
\hline
HJD\tablenotemark{a}-2400000 & RV (km~s$^{-1}$)\tablenotemark{b} \\
\hline
\endhead
58852.5632 & -22.5 \\
58853.5634 & -23.5 \\
58854.5636 & -24.1 \\
58855.5634 & -24.4 \\
58856.5640 & -22.4 \\
58857.5637 & -18.5 \\
58906.0286 & 4.1 \\
58914.0194 & -25.4 \\
58916.0138 & -22.5 \\
58930.9990 & 18.2 \\
58964.8884 & 7.2 \\
58965.8807 & 2.8 \\
58966.8795 & -0.8 \\
59020.7152 & 17.0 \\
59021.9444 & 14.5 \\
59022.9445 & 13.0 \\
59023.9546 & 10.7 \\
59024.9547 & 6.8 \\
59026.9548 & -0.9 \\
59028.9549 & -10.6 \\
59029.9550 & -14.7 \\
59100.6590  &  -5.6 \\                   
59103.6174  &   5.2 \\                   
59108.6144  &  15.9 \\                   
59109.6126  &  16.6 \\               
59110.6084  &  17.3 \\              
59112.6071  &  14.0 \\               
59113.6036  &  11.4 \\              
59115.7194  &   3.4 \\                
59116.6047  &  -0.2 \\               
59117.6367  &  -4.9 \\               
59118.7625  & -10.1 \\               
59119.6020  & -13.4 \\               
59120.6007  & -18.4 \\               
59121.6000  & -21.6 \\               
59122.5998  & -24.2 \\               
59123.5987  & -25.5 \\               
59124.5981  & -25.5 \\               
59125.5968  & -24.4 \\                
59126.5970  & -23.0 \\              
59127.5931  & -19.3 \\               
59128.5919  & -14.9 \\               
59129.5915  & -12.2 \\               
59130.5907  &  -7.4 \\               
59131.5903  &  -2.7 \\                
59132.5898  &   2.0 \\                
59133.5887  &   5.6 \\                
59134.5881  &   9.0 \\                 
59135.5869  &  12.3 \\               
59136.5867  &  13.6 \\              
59137.5860  &  15.9 \\               
59138.5852  &  16.4 \\               
59139.5838  &  17.0 \\                
59140.5839  &  16.6 \\              
59141.5830  &  15.4 \\
\hline
\end{longtable}
\tablenotetext{a}{HJD = Heliocentric Julian Date}
\tablenotetext{b}{Errors on RV are 0.7 km~s$^{-1}$.}

\begin{table}[H]
\centering
\caption{$\alpha$ Del A-Bb Astrometry}
\label{table:astrometry-hd196867b}
\begin{tabular}{lcccccc}
\hline
\colhead{UT Date} & \colhead{MJD} & \colhead{sep (mas)}   & \colhead{P.A. ($^\circ$)} & \colhead{error major axis (mas)} & \colhead{error minor axis (mas)} & \colhead{error ellipse P.A. ($^\circ$)} \\
\hline
2018Jul19 & 58318.3213 & 95.554 & 332.891 & 0.0858 & 0.0332 & 310.051 \\
2018Aug21 & 58351.2246 & 99.662 & 328.813 & 0.1008 & 0.0479 & 309.945 \\
2018Sep19 & 58380.2076 & 101.635 & 325.218 & 0.0709 & 0.0502 & 285.112 \\
2019Jun03 & 58637.4281 & 121.028 & 301.098 & 0.0264 & 0.0168 & 311.244 \\
2019Jul29 & 58693.2785 & 126.583 & 297.173 & 0.0372 & 0.027 & 317.253 \\
2019Jul30 & 58694.2693 & 126.388 & 297.023 & 0.0287 & 0.0197 & 344.943 \\
2019Jul31 & 58695.2677 & 126.193 & 296.829 & 0.0862 & 0.0523 & 313.489 \\
2019Aug01 & 58696.3156 & 126.019 & 296.595 & 0.0287 & 0.0204 & 315.051 \\
2019Aug06 & 58701.2788 & 126.233 & 295.304 & 0.0213 & 0.0171 & 315.130 \\
2019Aug08 & 58703.2806 & 127.012 & 294.884 & 0.0382 & 0.0347 & 299.196 \\
2019Nov12 & 58799.0774 & 137.151 & 288.352 & 0.1059 & 0.0802 & 33.258 \\
\hline
\end{tabular}
\end{table}

\begin{table}[H]
\centering
\caption{$\alpha$ Del: Best fit orbital elements}
\label{orbitelements:alpdel}
\begin{tabular}{lccc}
\hline
\colhead{} & \colhead{Outer Orbit} & \colhead{Inner Orbit - RV} & \colhead{Inner Orbit - Combined} \\
\hline
$P$ (d) & $6175.3 \pm 3.2$ & $29.979 \pm 0.011$ & $29.9873 \pm 0.0021$ \\
$T$ (MJD) & $57988.7 \pm 1.1$ & $58672.73 \pm 0.43$ & $58672.84 \pm 0.11$ \\
$e$ & $0.4615 \pm 0.0016$ & $0.0665 \pm 0.0058$ & $0.0761 \pm 0.0012$ \\
$\omega$ $(^\circ)$ & $91.73 \pm 1.09$ & $162.6 \pm 5.2$ & $166.43 \pm 1.27$ \\
$\Omega$ $(^\circ)$ & $120.62 \pm 1.19$ & -- & $359.97 \pm 0.65$ \\
$i$ $(^\circ)$ & $161.01  \pm 0.30$ & -- & $22.11  \pm 0.61$ \\
$a$ (mas)\tablenotemark{a} & $158.09 \pm 0.13$ & -- & $3.587 \pm 0.010$ \\
$M_{\rm{Ba}}/M_{\rm{Bb}}$ & -- & -- & $1.220 \pm 0.010$ \\
$K_{\rm{Bb}}$ (km~s$^{-1}$) & -- & $21.12 \pm 0.12$ & $21.19 \pm 0.18$ \\
$\gamma$ (km~s$^{-1}$) & -- & $-2.646 \pm 0.091$ & $-2.56 \pm 0.13$ \\
$T_{\rm{rv,max}}$ (MJD) & -- & -- & $58659.16 \pm 0.04$ \\
\hline
\end{tabular}
\tablenotetext{a}{$\pm0.25\%$ from absolute wavelength precision}
\end{table}

\begin{figure}[H]
\centering
\includegraphics[width=7in]{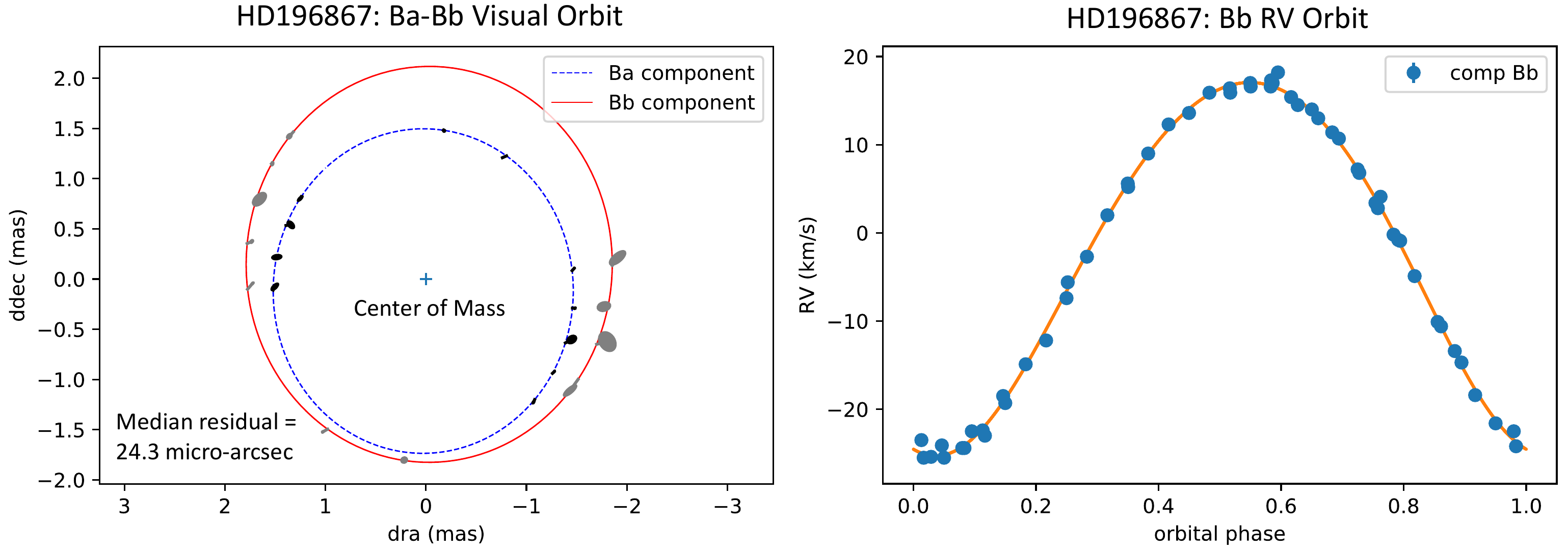}
\caption{We subtract the outer orbit of $\alpha$ Del to reveal the photocenter motion of the Ba component around the center-of-mass of the newly detected Ba-Bb system. Since we detect flux from the new Bb component, we can also show the Bb position. The motion of Ba+Bb about their center-of-mass allows us to work out the mass ratio of this inner pair, which in turn leads to measurements of all three masses in the system.}
\label{hd196867_inner_orbit}
\end{figure}

\begin{figure}[H]
\centering
\includegraphics[width=7in]{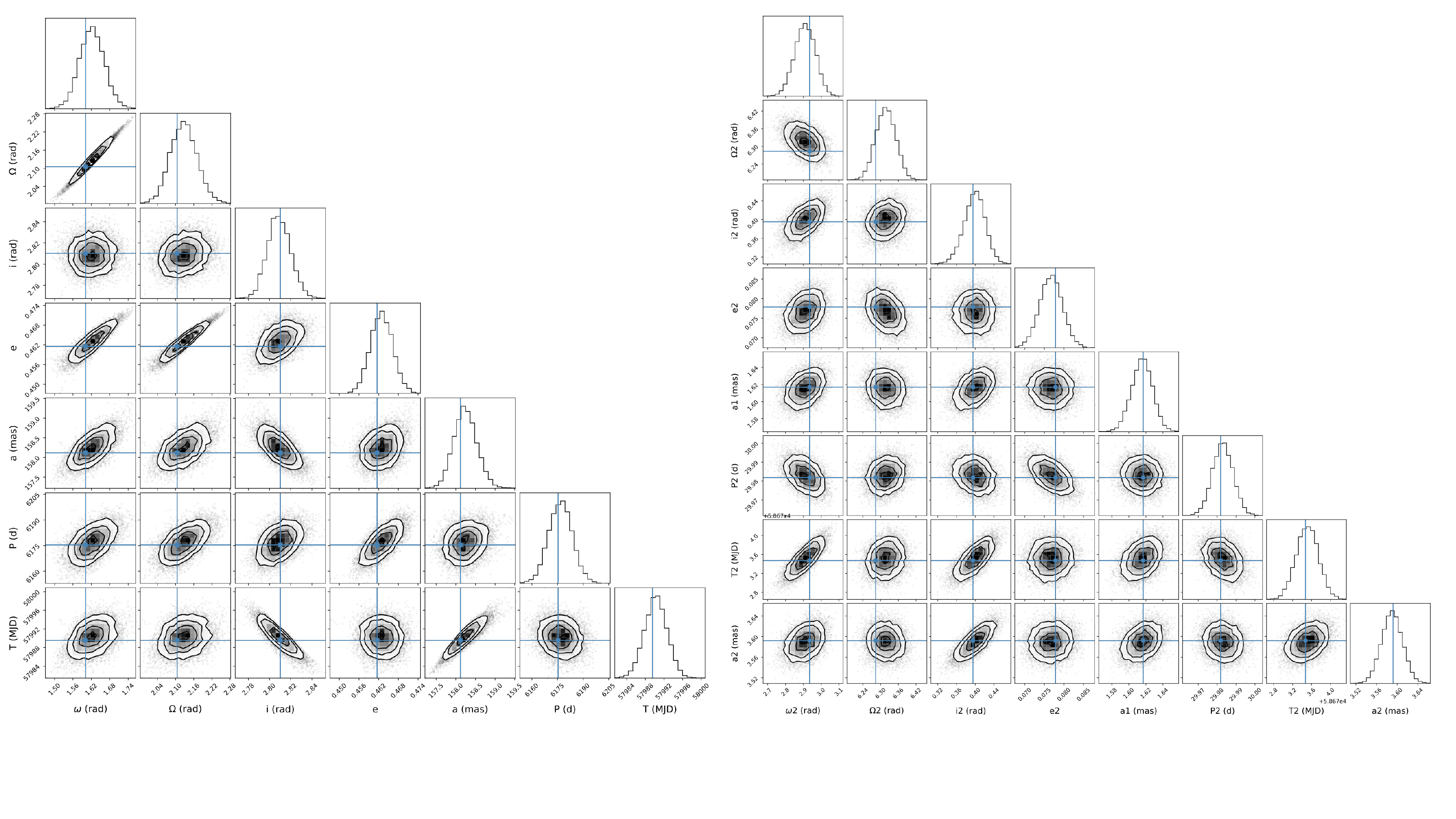}
\caption{We show corner plots for the inner (left) and outer (right) orbital elements of $\alpha$ Del. The inner orbit is particularly well constrained in this case, with little correlation among orbital parameters. }
\label{alpdel_corner}
\end{figure}


\begin{table}[H]
\centering
\caption{$\alpha$ Del: Physical Properties}
\label{physicalelements:alpdel}
\begin{tabular}{lc}
\hline
\hline
$f_{\rm{A}} / f_{\rm{Ba}}$ (H-band) & $10.086 \pm 0.038$\\
$f_{\rm{A}} / f_{\rm{Bb}}$ (H-band) & $14.261 \pm 0.069$ \\
distance (pc) & $78.5 \pm 2.2$ \\
$M_{\rm{A}}$ ($M_{\odot}$) & $3.83 \pm 0.33$ \\
$M_{\rm{Ba}}$ ($M_{\odot}$) & $1.82 \pm 0.15$ \\
$M_{\rm{Bb}}$ ($M_{\odot}$) & $1.49 \pm 0.12$ \\
$a_{\rm{outer}}$ (au) & $12.7\pm 0.4$ \\
$a_{\rm{inner}}$ (au) & $0.281 \pm 0.008$ \\
\hline
\end{tabular}
\end{table}

\subsection{$\nu$ Gem}
\label{sec:nugem}

$\nu$ Gem (HD 45542, HIP 30883, WDSJ06290+2013) is a bright ($V$=4.14, $H$=4.43) binary with a B-type primary and a Be star for the visual secondary at $\sim$0.1". This B+Be pair has a visual magnitude difference of about 1, with a period of about 19 years reported in the ORB6 catalog. \citet{jarad1989} discovered a third star in this system with a period of 40-days, though the authors pointed out that they are not confident in this period value. \citet{rivinius2006} confirmed this inner spectroscopic binary from radial velocity variations seen between 1994-2003, with a best-fit period of 53.72 days. These authors note that the circumstellar Balmer emission and shell lines do not take part in the 53.75-day period, concluding that the shell star is component B. The hierarchy of this system adopted from \citet{rivinius2006} is Aa+Ab for the inner pair, with the B component of the wide AB orbit being the Be star (note that this notation differs from the WDS catalog, which denotes the wide pair presented here as Aa--Ab with a B component at $>$100 arcsec). We report the first astrometric detection of the inner Aa+Ab orbit as a ``wobble" to the outer binary motion, with a period in agreement with \citet{rivinius2006}.

Recovering the flux from the inner component is difficult for this system since the signal is dominated by the bright wide binary, and our group delay tracker was noisier than usual for these datasets. 
When coupled with poor calibration, this makes a detection of a close ($<$1 mas in some epochs) companion difficult. We were able to confidently detect the inner companion in two of our MIRC-X epochs near apastron. To add additional data points to the inner visual orbit, we used 2 old MIRC datasets from 2015 and 2016. These datasets were taken in the lower resolution PRISM mode (R$\sim$42), meaning that the wide component is outside the interferometric FOV on most baselines. The stable fringe tracking on these nights and traditional TARGET-CALIBRATOR calibration sequence make it easier to find the inner component. The MIRC datasets were reduced with the standard MIRC data pipeline in IDL described in previous MIRC papers (e.g. \citealt{monnier2012}), with a coherent integration time of 75ms. For 2015Nov23 we used calibrators 71 Ori ($UD=0.597\pm0.021$, \citealt{schaefer2010}), HD23183 ($UD=0.854\pm0.059$, \citealt{lafrasse2010}), HD37329 ($UD=0.709\pm0.049$, \citealt{lafrasse2010}), and tet Gem ($UD=0.796\pm0.022$, \citealt{schaefer2010}). On 2016Nov14 we used calibrators 71 Ori, HD23183, and HD37329. We scaled the wavelengths produced by the pipeline by a factor of 1.004 to bring to an absolute scale as suggested by \citet{monnier2012}. For these old MIRC datasets, we used an IDL binary grid search routine\footnote{\url{http://www.chara.gsu.edu/analysis-software/binary-grid-search}} modified to fit the system as a triple \citep{schaefer2016}. The component diameters were fixed as point sources. Although the wide binary was outside the interferometric field-of-view on many baselines, we still fit for this wide component in our MIRC data. However, since the old MIRC epochs do not have our etalon wavelength calibration data, we are unable to bring these data to the same scale as our more recent MIRC-X nights. We assign the MIRC epochs errors of 0.4 mas (consistent with the largest etalon scale factor correction of 0.4\% = 0.4 mas for a 100 mas binary). At separations of a few milli-arcseconds for the inner pair, any wavelength scalings at the 1e-3 level have no effect on the measured separation within reported error bars. The MIRC positions are reported along with the MIRC-X astrometry in Tables \ref{table:astrometry-hd45542} and \ref{table:astrometry-hd45542_2}.

In order to measure orbital parallax, we initially combined our astrometric data with the single-line RV provided by \citet{rivinius2006}. Our joint visual+RV orbit yielded a ``wobble" semi-major axis of 1.47 mas, inclination of 81 degrees for the inner orbit, inner period of near 54 days, and a RV semi-amplitude of 35 km~s$^{-1}$ for the primary. These values lead to a distance of 125 pc, implying masses of the system much too low to be consistent with B-type stars. 
With an outer semi-major and period of 83 mas and 6974 days, the total mass in the system is only $\sim$3 $M_{\odot}$ at a distance of 125 pc. In private communication with the authors we learned that there are new RV data for this system, and that an upcoming paper will update the RV orbit. The published RVs in \citep{rivinius2006} were measured assuming an SB1 orbit, when in fact the spectral lines are a blend of the two inner components, leading to inaccurate RVs. Klement et al. (in prep) will include a full analysis of the 3 stars in this system, combining our astrometric new data with their updated RV analysis of all three components. Hence we present the visual orbit of $\nu$ Gem, without joint fitting of the problematic RVs. In Table \ref{physicalelements:nugem} we assume the Hipparcos distance of $167\pm8$ pc \citep{vanLeeuwen2007} when computing masses. 

When fitting to the WDS + ARMADA data for $\nu$ Gem, we find that the best-fit orbital parameters change between fitting the inner ``wobble" alone vs. fitting the inner visual orbit. As can be seen in Table \ref{orbitelements-nugem}, our ``wobble" orbit prefers a more eccentric solution. Our median residual to the outer binary + inner wobble motion for this orbit is 14.8 $\mu$arcsec. When coupling this motion with the position of the Ab component, however, our median residual increases to 40 $\mu$arcsec and the solution becomes less eccentric. This added residual in the wobble could imply motion from an additional companion in the system (which can be fit out by driving up eccentricity). In fact, there is accumulating evidence that nearly all Be stars are close binaries (e.g. \citealt{klement2017,klement2019}). The Hipparcos distance of this system leads to a mass sum which is already too low for B-type stars, implying that additional companions would need to be low mass -- although new RV data might update this distance. Recent work suggests that the companions to Be stars are generally low-mass, stripped down cores of former mass donors \citep{wang2017, bodensteiner2020}. Though additional companions are still possible, more epochs are needed to confirm the orbit of a fourth body in this system. It is also possible that the residuals can be explained by resolved time-varying structures in the Be disk. Since the ``wobble" depends on the measurement from star Aa -- B, this structure would add residual motion to our measured ``wobble" orbit. 

Our measured masses for the three components of this system are $2.7\pm0.4$ M$_{\odot}$, $2.5\pm0.4$ M$_{\odot}$, and $1.8\pm0.4$ M$_{\odot}$ for components Aa, Ab, and B respectively. The higher error bars come from the high error on parallax from Hipparcos. These masses assume that the Hipparcos distance for the system is accurate, which often is not the case for close binaries and triples. \citet{rivinius2006} listed B6III and B8IIIe as the preferred spectral types for the primary and wide component, though the spectral classification of Be stars in general is very uncertain due to weak photospheric lines. Classifications for this system in literature have ranged from mid-to-late B-type stars. Our reported masses are too low for these spectral classifications, likely implying that the distance is too low. Gaia DR2 lists an even lower distance of $155\pm14$ pc \citep{gaia2018}, though the lack of a multiple star solution makes this measurement untrustworthy. Klement et al. (in prep) will be able to measure an orbital parallax by combining their updated RV analysis with the astrometry presented in this work, which will lead to a more accurate measurement of masses. 

The orbital elements of the outer pair of $\nu$ Gem have higher errors bars when compared to $\alpha$ Del. This is due to worse coverage of this $\sim$19 year period. Figure~\ref{nugem_corner} shows posterior distributions of the outer and inner orbital elements. Again, there are no significant correlations between the parameters of the inner orbit with the parameters of the outer orbit -- so we split them up for clarity. Looking at the outer elements, one can see many correlations between parameters due to incomplete coverage of this outer orbit. Better coverage of this wide orbit may alter the outer orbital elements, particularly constraining semi-major axis and inclination more confidently. These elements are important for constraining the total mass in the system, and better coverage of this wide orbit would lead to better mass constraints.

\begin{table}[H]
\centering
\caption{$\nu$ Gem Aa-B Astrometry}
\label{table:astrometry-hd45542}
\begin{tabular}{lcccccc}
\hline
\colhead{UT Date} & \colhead{MJD} & \colhead{sep (mas)}   & \colhead{P.A. ($^\circ$)} & \colhead{error major axis (mas)} & \colhead{error minor axis (mas)} & \colhead{error ellipse P.A. ($^\circ$)} \\
\hline
2015Nov23\tablenotemark{a} & 57349.374 & 77.809 & 114.589 & -- & -- & -- \\
2016Nov14\tablenotemark{a} & 57706.437 & 87.121 & 118.536 & -- & -- & -- \\
2017Sep28 & 58024.5609 & 93.148 & 121.903 & 0.053 & 0.029 & 90.00 \\
2017Sep30 & 58026.5487 & 93.201 & 121.905 & 0.041 & 0.018 & 121.67 \\
2018Sep20 & 58381.5053 & 95.478 & 125.440 & 0.159 & 0.029 & 89.93 \\
2018Nov21 & 58443.462 & 93.947 & 125.936 & 0.093 & 0.032 & 42.83 \\
2018Dec04 & 58456.3938 & 92.303 & 125.955 & 0.114 & 0.049 & 137.61 \\
2019Sep08 & 58734.5365 & 88.279 & 128.610 & 0.059 & 0.034 & 348.69 \\
2019Oct13 & 58769.5102 & 87.900 & 129.244 & 0.028 & 0.026 & 25.13 \\
2019Nov11 & 58798.4763 & 88.462 & 129.284 & 0.047 & 0.040 & 47.34 \\
\hline
\end{tabular}
\tablenotetext{a}{MIRC data. We assign 0.4 mas error bars to these data, since we do not have etalon calibration data for these older nights.}
\end{table}

\begin{table}[H]
\centering
\caption{$\nu$ Gem Aa-Ab Astrometry}
\label{table:astrometry-hd45542_2}
\begin{tabular}{lcccccc}
\hline
\colhead{UT Date} & \colhead{MJD} & \colhead{sep (mas)}   & \colhead{P.A. ($^\circ$)} & \colhead{error major axis (mas)} & \colhead{error minor axis (mas)} & \colhead{error ellipse P.A. ($^\circ$)} \\
\hline
2015Nov23\tablenotemark{a} & 57349.374 & 1.850 & 118.823 & 0.010 & 0.008 & 118.5216 \\
2016Nov14\tablenotemark{a} & 57706.437 & 2.965 & 312.417 & 0.012 & 0.008 & 141.4462 \\
2017Sep28 & 58024.562 & 2.774 & 307.268 & 0.055 & 0.028 & 286.86 \\
2017Sep30 & 58026.5487 & 2.920 & 309.670 & 0.043 & 0.023 & 305.08 \\
\hline
\end{tabular}
\tablenotetext{a}{MIRC data}
\end{table}

\begin{table}[H]
\centering
\caption{$\nu$ Gem: Best fit orbital elements}
\label{orbitelements-nugem}
\begin{tabular}{lcccc}
\hline
\colhead{} & \colhead{Outer Orbit} & \colhead{Inner Orbit: Wobble} & \colhead{Inner Orbit: Visual} & \colhead{Inner Orbit: Wobble + Visual} \\
\hline
$P$ (d) & $6985 \pm 18$ & $54.029 \pm 0.021$ & $53.742 \pm 0.088$ & $53.7276 \pm 0.0066$ \\
$T$ (MJD) & $55939 \pm 74$ & $58461.20 \pm 0.79$ & $58487 \pm 15$ & $58488.6 \pm 2.7$ \\
$e$ & $0.28 \pm 0.01$ & $0.198 \pm 0.032$ & $0.0373 \pm 0.022$ & $0.0303 \pm 0.004$ \\
$\omega$ $(^\circ)$\tablenotemark{a} & $233 \pm 3$ & $44.8 \pm 2.9$ & $13 \pm 98$ & $26 \pm 18$ \\
$\Omega$ $(^\circ)$\tablenotemark{a} & $120.19 \pm 0.28$ & $127.78 \pm 0.54$ & $131.21 \pm 0.96$ & $131.17 \pm 0.16$  \\
$i$ $(^\circ)$ & $75.92 \pm 0.15$ & $81.43 \pm 0.37$ & $79.38 \pm 1.8$ & $79.76 \pm 0.33$ \\
$a$ (mas)\tablenotemark{b} & $83.12 \pm 0.59$ & $1.478 \pm 0.017$\tablenotemark{c} & $2.892 \pm 0.049$ & $2.895 \pm 0.019$ \\
$M_{\rm{A_a}}/M_{\rm{A_b}}$ & -- & -- & -- & $1.117 \pm 0.035$ \\
$T_{\rm{rv,max}}$ (MJD) & -- & $58510.7 \pm 0.2$ & $58511.60\pm0.34$ & $58511.34 \pm 0.09$ \\
\hline
\end{tabular}
\tablenotetext{a}{Since we do not include RV, there is a 180$^{\circ}$ degeneracy. We report the $\Omega<180^{\circ}$.}
\tablenotetext{b}{$\pm0.25\%$ from absolute wavelength precision}
\tablenotetext{c}{``wobble" semi-major}
\end{table}

\begin{figure}[H]
\centering
\includegraphics[width=6in]{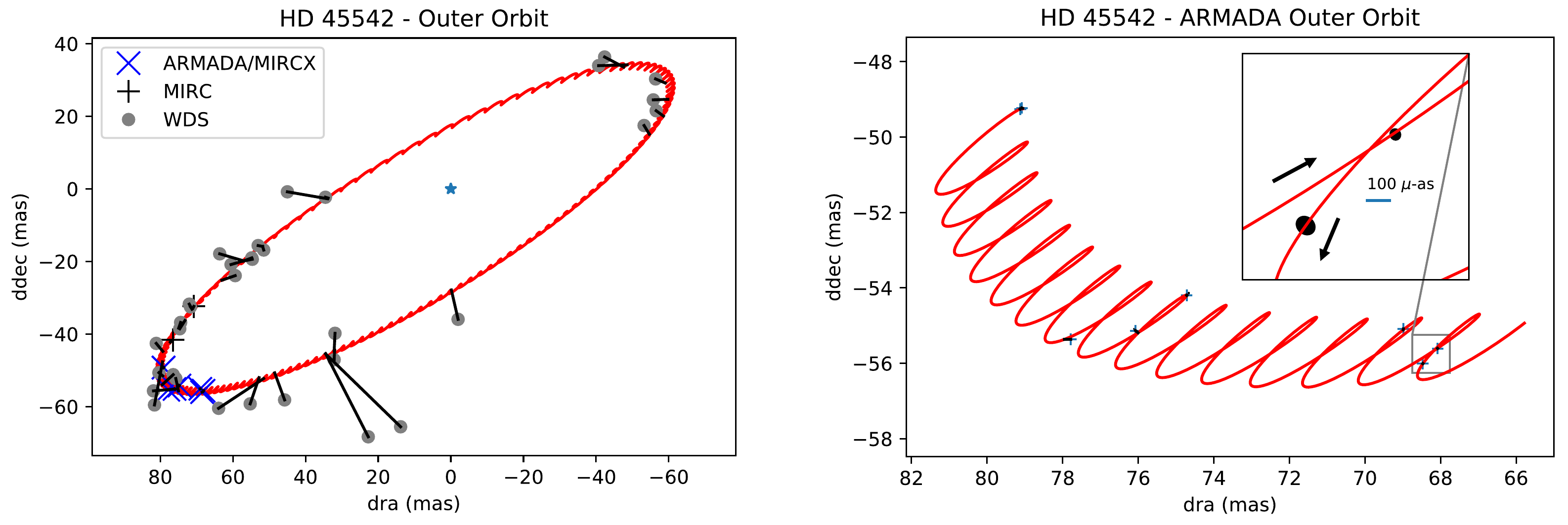}
\caption{We detected the inner tertiary component to the system of B-type binary $\nu$ Gem. We show the best fit to the ARMADA+WDS data for the outer binary pair - we zoom in to the ARMADA portion of the orbit to show that there is clearly an additional ``wobble" to this binary motion.}
\label{nugem_orbit}
\end{figure}

\begin{figure}[H]
\centering
\includegraphics[width=7in]{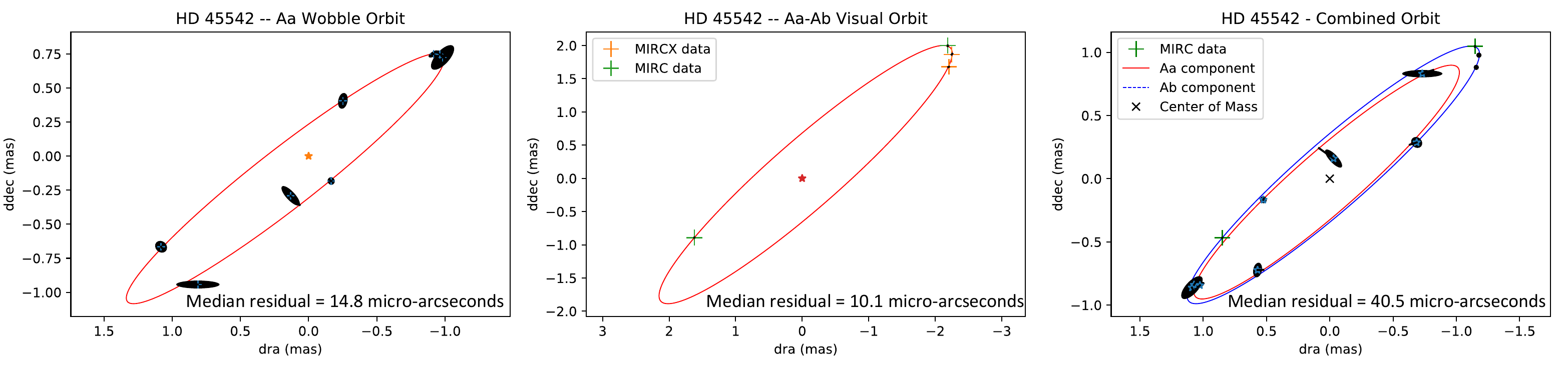}
\caption{(Left) Subtracting out the binary motion of $\nu$ Gem, we show the fit results of the inner wobble of the primary star induced by the $\sim$54 day companion. Here we are fitting only to the outer binary motion + wobble. (Center) We also detect flux from the Ab component in four MIRC-X epochs, and in two epochs of MIRC data. We plot the best fit orbit when fitting only to this Aa-Ab pair. Visually, one can see that the best-fit orbit does not agree well with the orbit fit of the wobble motion alone. The wobble motion prefers a higher eccentricity orbit than that derived from the visual data alone, with a significantly larger period. (Right) We show the combined fit, when coupling the wobble motion to the visual orbit. The median residual of the orbit increases, due to the residuals from the wobble motion. This extra residual is potentially due to time-varying resolved structure from the Be star disk, or additional companions in the system. }
\label{nugem_orbit_inner}
\end{figure}

\begin{figure}[H]
\centering
\includegraphics[width=7in]{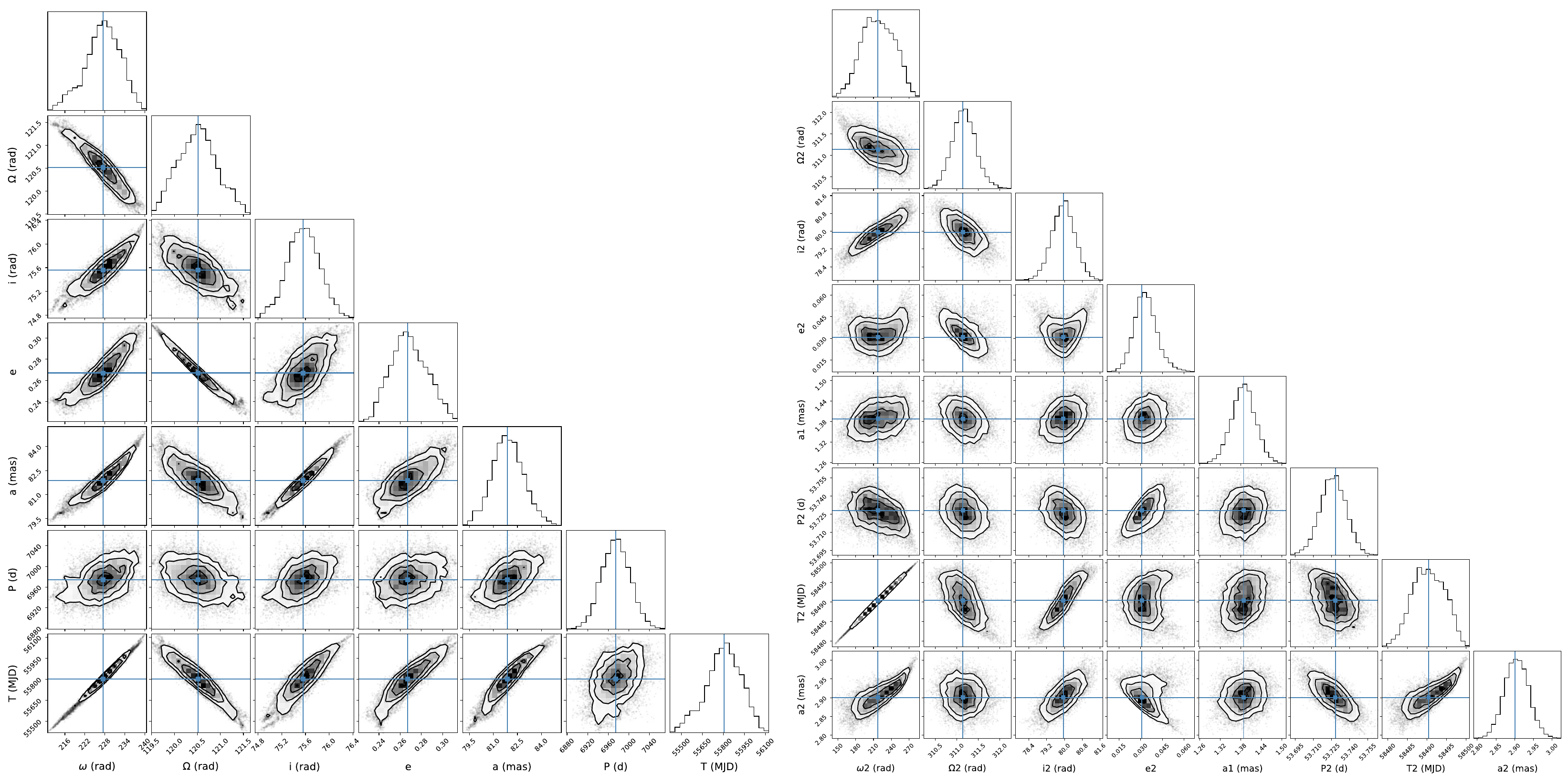}
\caption{We show corner plots for the inner (left) and outer (right) orbital elements of $\nu$ Gem. Since the outer 19-year period is not very well constrained, many of the orbital parameters show correlations with each other. Better coverage of the outer orbit will lead to tighter constraints. The inner orbit parameters show fewer correlations, with the strongest being between the angle and time of passage through periastron -- expected for a near circular orbit. }
\label{nugem_corner}
\end{figure}


\begin{table}[H]
\centering
\caption{$\nu$ Gem: Physical Properties}
\label{physicalelements:nugem}
\begin{tabular}{lc}
\hline
\hline
$f_{\rm{Aa}} / f_{\rm{Ab}}$ (H-band) & $1.95 \pm 0.39$ \\
$f_{\rm{Aa}} / f_{\rm{B}}$ (H-band) & $1.286 \pm 0.041$ \\
distance (pc) & $167 \pm 8$\tablenotemark{a} \\
$M_{\rm{Aa}}$ ($M_{\odot}$) & $2.7 \pm 0.4$ \\
$M_{\rm{Ab}}$ ($M_{\odot}$) & $2.5 \pm 0.4$ \\
$M_{\rm{B}}$ ($M_{\odot}$) & $1.8 \pm 0.4$ \\
$a_{\rm{outer}}$ (au) & $13.7 \pm 0.7$ \\
$a_{\rm{inner}}$ (au) & $0.48 \pm 0.02$ \\
\hline
\end{tabular}
\tablenotetext{a}{The Hipparcos distance is assumed \citep{vanLeeuwen2007}. This affects the measurement of masses and physical semi-major axes. Klement et al. (in prep) will use our data to obtain a better measurement of orbital parallax.}
\end{table}

\subsection{Implications for formation}
\label{sec:formation}
These triple systems exemplify the power provided by complete orbital information, particulary in the case of the well-constrained $\alpha$ Del orbit. Our analysis shows that the Ba-Bb binary orbit of $\alpha$ Del is retrograde with respect to its orbit about the primary, with a mutual inclination of $159.8\pm0.5^{\circ}$. Though not unique (see e.g. \citet{Tokovinin:2020}, retrograde orbits are atypical. Unlike many known high inclination triples $\alpha$ Del is not subject to Kozai-Lidov oscillations \citep{Lidov:1962,Kozai:1962}. With near equal mass ratio and an eccentric outer orbit, the system stretches the limits of current secular theory \citep{Naoz:2016}. Thus we verify via n-body simulations using Rebound \citep{Rein:2012} that the full orbital solution is stable against secular instability, and merely undergoes nodal precession and small amplitude eccentricity oscillations. Notably the current oscillating orbital elements are very close to the median values found over 1000s of orbits.

That the system is retrograde, stable against large inclination excitation, and compact, suggests a possibly violent dynamical interaction in the past. Formation within the same disk via fragmentation is strongly disfavored given the mutual inclination and short orbital periods. \citep{Kratter:2010,Hall:2017}. Formation via turbulent fragmentation on large scales (or a combination of both modes of fragmentation) requires substantial migration either via gas-drag or dynamical interactions; the former case would favor smaller mutual inclinations \citep{Lee:2019}. In the latter case, the tightness of the orbits favors the ejection of one or more previously bound objects to serve as the requisite sink of energy and angular momentum.

An expected consequence of dynamical instability is the misalignment of orbital and stellar spin vectors. While the obliquities of the three components of $\alpha$ Del cannot be measured directly, we can infer from measured rotational velocities that alignment is not favored for two of the three components. For Ba, the measured $v\sin i \approx 60$km/s is consistent with typical spin rates for mid-A stars if it is aligned with either the Ba-Bb orbit ($v\approx 160$ km/s at $22^\circ$)  or the BaBb-A orbit ($v\approx 180$ km/s at $161^\circ$) \citep{Zorec:2012}. However, with $v \sin i \approx 7$ km/s, the Bb component would have to be a very slow rotator ($v\approx 20$ km/s) if it shared Ba's obliquity, unusual for a star above the Kraft break.  Similarly, the primary component's $v\sin i \approx 145$km/s is hard, though not impossible, to reconcile with an inclination of $161^\circ$: a nearly pole on orbit with $v\approx 445$ km/s implies that the star is rotating near breakup velocity. With our current data, it is not yet possible to rule out the expected overluminosity or distortion consistent with such rapid rotation. 

Without the inclusion of RV information, $\nu$ Gem is harder to constrain from a formation history perspective due to the ambiguity of the position angles of the ascending node ($\Omega$) for the inner and outer orbits. In contrast with $\alpha$ Del, our $\nu$ Gem data is consistent with nearly co-planar orbits -- though RV data will break the degeneracy. Co-planar orbits often have a formation history with sequential epochs of disk fragmentation followed by migration \citep{TokovininMoe:2020}. Disk instability becomes increasingly likely at higher stellar masses \citep{Kratter:2008}. The upcoming RV analysis for $\nu$ Gem, including measurements of $v\sin i$ for some components, will increase the clarity of mutual inclinations and potential stellar obliquities in this system (Klement et al., in prep).

\section{Summary and Future Work}
\label{sec:conclusion}

We started the ARMADA survey with the MIRC-X instrument at the CHARA array with the goal of detecting giant planets on $\sim$au orbits and previously unseen low-mass companions orbiting individual stars of binary systems. In this paper, we introduced the observational methods and calibration scheme which can bring us to $\sim$10 micro-arcsecond precision for our differential binary orbits. Our newly implemented etalon module for precise wavelength calibration was demonstrated to improve systematic errors in binary separation on the triple star test system $\kappa$ Peg. We achieved a median precision level of 6.3 micro-arcseconds for data taken after our 2018Sep optics upgrade of MIRC-X, and demonstrated a vast improvement in astrometric performance compared to previous similar interferometric surveys. We presented the updated RV+visual orbit of this triple star system. 

Though our astrometric precision is about a factor of 10 better than previous work, there is still potential to do better in approaching the fundamental limits of interferometric observations. \citet{ireland2018} predict sub-microarcsecond atmospheric and shot-noise limited precision for the $\sim$100 milli-arcsecond binaries presented in this paper. Hence another factor of $\sim$10 improvement is theoretically possible, but would require calibration of pupil registration at the $<$1 mm level projected onto primary telescope mirror space. Future ARMADA papers will included more detailed studies in the factors currently limiting precision. 

Confident detections of giant planets will take a longer time-baseline and higher number of epochs, but are beginning to see the ``wobble" signature from the gravitational tug of previously unseen short-period tertiary companions. Since these companions are stellar, we often detect the flux from the stars as well and measure the mass ratio of the inner pair. Combined with single-line RV data, we are able to measure orbital parallax along with the masses of all three components in these systems. We made the first discovery and measured the orbital elements of a 30-day companion to the B-type binary $\alpha$ Del, which includes our new MIRC-X/ARMADA data as well as new RV points from the Fairborn Observatory. We also detected for the first time the inner visual orbit of B-type triple star $\nu$ Gem. This system is of particular importance, since it will become part of an ongoing study to understand multiplicity in Be star systems. Our detection of the inner orbit is crucial for solving the physical parameters of the system, and an upcoming paper by Klement et al. (in prep) will characterize this system by including new RV data to this inner and outer orbit. 

Our data is consistent with $\alpha$ Del being born of a violent dynamical instability / ejection event, while being more uncertain for $\nu$ Gem without RV data. The rich dynamics of these systems demonstrate the power of full orbit solutions for revealing the origins of triple systems (both planetary and stellar). Future discoveries that contain, for example, more compact inner binaries may also prove useful for constraining basic tidal evolution models.

As is evident from these first three systems, our ARMADA survey will be extremely efficient at detecting $<$1 au stellar companions in wide binary systems. In future work, we plan to publish our full list of newly detected triple systems with the ARMADA survey. When combined with RV of the outer and inner pairs, we will be able to fully characterize these orbits, including masses (which can be measured using the methods of this paper), along with mutual inclinations (if combined with RV), and evolution history on an HR diagram. 

The residuals to our best-fit orbits are also promising for detecting $\sim$au brown dwarfs and giant planets in interesting regimes. We target mainly A/B-type stars for ARMADA, where the $\sim$au giant planet occurrence is difficult to measure due to weak and broad spectral lines of these stars. Binary systems themselves are crucial regimes for searching for planets, as circumstellar planets in close binaries are difficult to detect. Finding planets in binary systems provides clues to the formation timescales and channels for giant planets. With more epochs, our ARMADA survey should be able to detect giant planets in these regimes. 

\acknowledgments
T.G. and J.D.M. acknowledge support from NASA-NNX16AD43G, and from NSF-AST2009489. T.G. acknowledges support from Michigan Space Grant Consortium, NASA grant NNX15AJ20H. S.K., N.A., and C.L.D acknowledge support from ERC Starting Grant No. 639889. A.L. acknowledges support from studentship No. 630008203 by the UK Science and Technology Facilities Council.
This work is based upon observations obtained with the Georgia State University Center for High Angular Resolution Astronomy Array at Mount Wilson Observatory.  The CHARA Array is supported by the National Science Foundation under Grant No. AST-1636624 and AST-1715788.  Institutional support has been provided from the GSU College of Arts and Sciences and the GSU Office of the Vice President for Research and Economic Development. MIRC-X received funding from the European Research Council (ERC) under the European Union's Horizon 2020 research and innovation programme (Grant No. 639889).
This research has made use of the Jean-Marie Mariotti Center SearchCal service\footnote{available at \url{http://www.jmmc.fr/searchcal_page.htm}}. We thank Nuria Calvet for supporting funds in the development of our etalon wavelength calibration module. This research has made use of the Washington Double Star Catalog maintained at the U.S. Naval Observatory.

%
%



\vspace{5mm}
\facilities{CHARA, Fairborn Observatory}
\software{lmfit, astropy, numpy, emcee}



\bibliographystyle{aasjournal}
\bibliography{references}

\listofchanges

\end{document}